\documentclass[
reprint,
superscriptaddress,
amsmath,
amssymb,
aps,
pra,
showkeys
]{revtex4-2}

\usepackage[utf8]{inputenc}
\usepackage{xcolor, ulem, booktabs, multirow, amscd, graphicx, epstopdf, appendix, hyperref, color, setspace, tensor, url, verbatim, float, mathrsfs, amsmath, amssymb, physics, placeins, setspace, natbib, dcolumn, bm}

\newcommand{\hide}[1]{}

\newcommand{\eq}[1]{Eq.\,\,(\ref{#1})}

\newcommand{\eqp}[1]{Eqs.\,\,(\ref{#1})}

\newcommand{\fig}[1]{Figure\,\,\ref{#1}}

\newcommand{\evs}[2]{\expval{\hat{\sigma}_{#1#2}}}
\newcommand{\eva}{\expval{\hat{a}}}
\newcommand{\evad}{\expval{\hat{a}^{\dagger}}}
\newcommand{\evada}{\expval{\hat{a}^{\dagger}\hat{a}}}
\newcommand{\mc}[1]{\mathcal{#1}}
\newcommand{\mb}[1]{\mathbf{#1}}

\newcommand{\Emcal}{\mathcal{E}}

\newcommand{\trn}[2]{\ket{#1}-\ket{#2}}
\newcommand{\tens}[1]{\bar{#1}}
\newcommand{\mathsym}[1]{{}}
\newcommand{\unicode}[1]{{}}

\graphicspath{ {images/} }

\begin{document}

\preprint{}

\title{Effect of photon propagation on a zero refractive index medium}

\author{Robert A. McCutcheon}
\affiliation{%
 Physics Department, University of Connecticut, Storrs, CT 06269}
\author{Stefan Ostermann}%
\affiliation{%
 Physics Department, Harvard University, Cambridge, MA 02138}%
\author{Susanne F. Yelin}
\affiliation{%
 Physics Department, Harvard University, Cambridge, MA 02138}%
 \affiliation{%
 Physics Department, University of Connecticut, Storrs, CT 06269}

\date{\today}

\begin{abstract}
We present a model describing the transmission of light through atomic media with a vanishing index of refraction. Zero index materials are of particular interest as the infinite phase velocity of light within the material offers the potential to manipulate electromagnetic waves to mediate dipole-dipole interactions over extended distances. We focus on the preparation of zero-index conditions based on atomic coherence using two distinct atomic media as exemplary of generic zero-index materials. We establish a model based on the Maxwell-Bloch equations to describe the propagation of a light pulse through these media. To investigate the sustainability of the zero index under minimal light conditions, we assume single-photon intensity of the propagating pulse. Specifically, we examine whether the spatial phase change of the photon remains zero as it traverses the medium. We employ a finite-element numerical approach to solve the coupled Maxwell-Bloch equations describing the photon propagation. Our results indicate that the presence of a photon within the medium will disrupt the zero-index state, thus disallowing the establishment of enhanced dipole-dipole interactions over large distances.
\end{abstract}


\maketitle

%
%
%
%
\section{Introduction}\label{sec:zi_intro}
\begin{figure}[t]
    \centering
    \includegraphics[trim={0cm 0cm 0cm 0cm},clip,width=\linewidth]{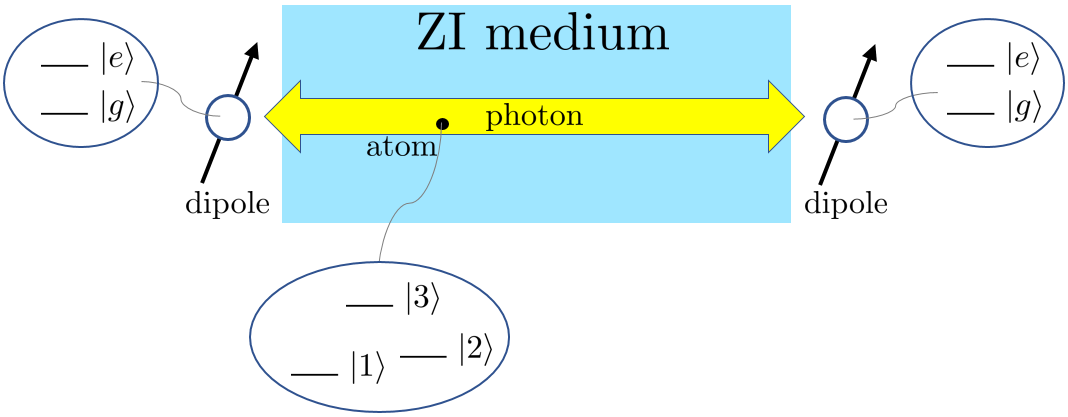}
	\caption{
 Schematic of dipole-dipole interaction across ZI medium. A field consisting of a single photon propagates through a medium consisting of atoms with a particular structure, prepared for ZI. The photon may interact with atoms along the way, possibly affecting its own refractive index, and couples dipoles located on either side of the medium. Other control fields may be present in order to prepare the ZI, but are not shown.}
     \label{fig:zi_dip-dip}
\end{figure}

The study of media with zero refractive index (ZI) has been growing recently because of their unique ability to modify the  propagation characteristics of electromagnetic waves \cite{liberal2016,liberal2017,mahmoud2017,alu2008,liberal2016-2,schilling2011,ziolkowski2004,horsley2021}.
This opens up various potential applications relevant for next-generation optical devices, ranging from unconventional wavefront manipulation to enhanced light-matter interactions \cite{suzuki2020,gadomsky2014,li2013,islam2015,bor2020,xu2014,luo2021}.
If the refractive index $n$ experienced by light of a particular frequency in some medium goes to zero, the wave number $k$ is expected to vanish, which means that the wavelength of the light and the phase velocity go to infinity. This implies that the phase of the light field does not vary in space throughout the medium. Also, full transmission of light through ZI channels of arbitrary shape is theoretically possible, which has been called ``supercoupling" \cite{liberal2016,liberal2017,mahmoud2017,alu2008,marcos2015,luo2018}.

Based on these unique properties it has been proposed that a ZI medium can be used to couple distant dipoles as strongly as if there was no medium (and no distance) between them. This is because the ZI medium behaves as an ``electromagnetic point" for light entering at one port and exiting at another; the light's response is insensitive to the ZI channel's size and shape and the orientation of the ports \cite{liberal2017,alu2008,sokhoyan2013}. It is as if the space is shrunk down, and the dipole-dipole coupling is as strong as if the medium were not there and they were closer together. Numerical simulations have verified the possibility of this phenomenon \cite{mahmoud2017,alu2008}. This effect relies mainly on the lack of spatial phase change of light in the medium; light entering is always in phase with light exiting, and light emitted by one dipole can be always in phase with light reaching the other across the ZI medium. 

The dipole-dipole interaction is mediated by photons as each dipole either emits or absorbs a single photon \cite{lehmberg1970, lehmberg1970-2, asenjo-garcia2017, asenjo-garcia2017-2, munro2018}. Distant coupling depends strongly on the lack of phase change of light throughout the ZI medium. However, it is possible that a ZI response in the medium is fragile; the presence of light may affect the response of the medium to it, which could ruin a prepared ZI response, and cause spatial phase change of the light. This brings us to the central question of this paper, which is: does a single photon ruin its own ZI in an atomic medium? 

There are two main approaches for obtaining a zero index (or other ``exotic" values of index, such as negative or very high): control of atomic media via external fields \cite{scully1991,kastel2007,kuznetsova2013}, and the design of metamaterials with desired electromagnetic properties \cite{schilling2011,pendry2006,smith2000,smith2004,suzuki2020,jiang2020}. We choose the former approach to serve as a general example of propagation under ZI, which is more straightforward to achieve, at least theoretically.

In this paper, we present a model for light propagation in atomic media to identify the characteristic absence of spatial phase change with ZI. First, we describe the general approach taken, which involves a pulse containing a single photon traveling through a gas consisting of atoms of a particular structure, prepared to give a ZI response for the light. Then we propose a medium consisting of three-level atoms and explain how a ZI can be prepared for it. Then we derive a Maxwell-Bloch model for the propagation of the light and the atomic response. We describe the numerical simulation technique used to solve the equations of motion and discuss its results. Finally, we follow these steps for a different medium which consists of five-level atoms.
%
%
%
%

\section{General Approach}\label{sec:zi_general_approach}
We suppose that a pulse of light is traveling in vacuum. The light encounters an ensemble of atoms, which are initially prepared to give a ZI response for the incoming light. Expectation values for the field and atomic variables will be calculated as functions of time and space using Maxwell-Bloch equations as the pulse proceeds through the medium. The index of refraction is also treated as a function of time and space, so that its instantaneous value affects the light, even as the light possibly changes the index through its own effect on the medium.

We arbitrarily choose the propagation to be in the $\hat{z}$ direction, with a general polarization in the transverse direction. Since the ZI response is experienced only by a particular field frequency, and for the sake of simplicity, we assume that the field is monochromatic with frequency $\nu$ and mode $\hat{a}$. The field has the complex wave number $k(z,t)=n(z,t)k_0$, where $k_0$ is the vacuum wave number. We use the form 
\begin{equation}\label{eq:zi_plane_wave_form}
    \mb{E}(z,t)=\frac{1}{2}\mathcal{E}(z,t)\hat{\epsilon}_E e^{in(z,t)k_0z-i\nu t}+c.c.
\end{equation} 
for the electric field. Using the expectation value of the quantum electric field operator, where $V$ is the quantization volume,
\begin{equation}
    \expval{\hat{\mb{E}}}(z,t) = \sqrt{\frac{\hbar\nu}{2\epsilon_0 V}}\eva(z,t) e^{in(z,t)k_0 z-i\nu t} + c.c.,
\end{equation}
the envelope $\mathcal{E}(z,t)$ is identified with $\sqrt{2\hbar\nu/\epsilon_0V}\eva (z,t)$. $\eva$ will be calculated as the field travels through the medium. In order to see whether the minimum amount of light affects its own refractive index as it interacts with the atoms, the envelope of the pulse is chosen so that the field contains one photon initially. We use a Gaussian distribution for $\evada(z,t)$, which travels along $z$ with speed $c$ before encountering the medium. We assume that the field is in a coherent state with a real eigenvalue, so that we have $\eva(z,t)=\sqrt{\evada(z,t)}$ for simplicity. The form of $\eva(z,t)$ in vacuum before encountering the medium therefore is
\begin{equation}\label{eq:zi_pulse}
    \eva(z\le0,t)=\left[\frac{1}{\sqrt{2\pi} \sigma_s}e^{\left(-z - z_i - c t\right)^2/(2\sigma_s^2)}\right]^{1/2} ,
\end{equation}
where $\sigma_s$ is the standard deviation of the distribution and $z_i$ is the initial position of its maximum. For the numerical simulations that follow, $\eva (0,0)$ is the value at the boundary to the medium that enters initially.

We suppose that there is an ensemble of atoms along $z$ from $z=0$ to $z=z_f$. The atoms are identical and do not interact with each other. Depending on the structure of the atoms and how the incoming field couples to them, we choose external control fields, the ensemble density, and other parameters such that the incoming field should experience ZI, at least initially.

We will only calculate the field from $z=0$ to $z=z_f$, when it is inside the medium. We do not consider impedance matching or reflection at the boundary; in effect, the pulse given by \eq{eq:zi_pulse} is the light that actually enters the medium to ensure that only one photon propagates through, after any reflection may have occurred in reality, although this may be avoided with ZI materials \cite{suzuki2020}.

Next, a set of Maxwell-Bloch equations are derived which are used to calculate the probe field's propagation and the atomic response \cite{scully2008, loudon2000}. From these results, we calculate the index of refraction and phase of the light. The instantaneous value of the index, which will be calculated differently depending on the atomic structure, may suggest whether there was any spatial phase change, but the effects of the overall index will ultimately be evident by looking at the phase of the field. Any spatial variation in the phase would suggest that the effective index is not zero. The effective index may contain contributions not represented by the instantaneous index $n(z,t)$ in \eq{eq:zi_plane_wave_form}.

The Maxwell equation for $\eva(z,t)$ determines how it changes in time and space through the medium. The Bloch equations are first-order differential equations in time for the atomic variables $\evs{i}{j}(z,t)$, where the operator $\hat{\sigma}_{ij}=\dyad{i}{j}$ for states $\ket{i},\ket{j}$, but through coupling to $\eva(z,t)$, these gain spatial dependence. We do not distinguish between individual atoms; rather the atomic expectation values are averages and treated as continuous in space. 

Throughout this paper, all expectation values, fields, field amplitudes, and the refractive index are functions of $z$ and $t$, but this explicit dependence may be dropped.

%
%
%
%
\section{Propagation Through Three-Level Atoms}

\subsection{Theoretical Model}\label{sec:zi_three-level_model}
\begin{figure}
    \centering
    \includegraphics[trim={0cm 0cm 0cm 0cm},clip,width=\linewidth]{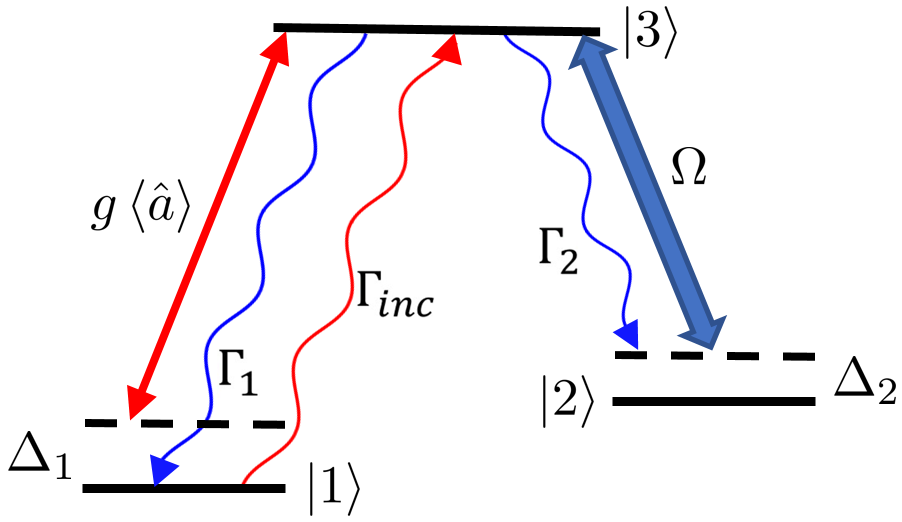}
	\caption{Three-level atom coupled to external fields, with incoherent pumping from $\trn{1}{3}$. $\Gamma_{inc}$ is the incoherent pumping rate, $\Gamma_1$ and $\Gamma_2$ are population decay rates, $g\eva$ is the effective Rabi frequency of the probe field, $\Omega$ is the Rabi frequency of the pump field, and $\Delta_1$ and $\Delta_2$ are detunings.}
     \label{fig:zi_three-level_atom}
\end{figure}

To address the requirement for a zero-index model that can equally incorporate the effect of an additional light field or a single photon, we initially consider a medium composed of the $\Lambda$-type atoms shown in \fig{fig:zi_three-level_atom} as the simplest possible case. With incoherent pumping on the transition coupled to the probe field, a ZI can be obtained. Although incoherent pumping gives a phenomenological pump rate that can come from variety of effects, its origin is irrelevant to our theoretical purpose and results.

The probe field couples to the $\trn{1}{3}$ transition, which is an electric dipole transition. This is treated quantum mechanically with dipole operator $\hat{\mb{d}}$, whose elements are assumed real, and coupling strength $g=\sqrt{\nu/(2\hbar\epsilon_0V)}d$. This transition has the decay rate $\Gamma_1$. The other transition is driven by a pump field, treated semi-classically with Rabi frequency $\Omega$, and has the decay rate $\Gamma_2$. The fields are detuned from the atomic resonances of their respective transitions by $\Delta_1$ and $\Delta_2$. The single-atom Hamiltonian for this system, under the rotating wave approximation (RWA) and in a rotating frame, is
\begin{equation}\label{eq:zi_three-level_hamiltonian}
\hat{H}/\hbar = \Delta_1\hat{\sigma}_{11} + \Delta_2\hat{\sigma}_{22} - \left( g\hat{a}\hat{\sigma}_{31} + \Omega\hat{\sigma}_{32} + \text{H.c.} \right) .
\end{equation}

There is a polarization induced by the probe field, which we take to be linear in the field: $\mb{P}=\epsilon_0\chi\mb{E}$. This depends on the coherence between levels $\ket{1}$ and $\ket{3}$ as follows:
\begin{equation}
    \mb{P}=N\ev{\hat{\mb{d}}}=N\mb{d}\left(\evs{1}{3}+\evs{3}{1}\right) ,
\end{equation}
which leads to 
\begin{equation}\label{eq:zi_three-level_susc}
    \chi = \frac{2Nd\evs{1}{3}}{\epsilon_0\mc{E}} = \frac{2 N_{atoms} g}{ck_0}\frac{\evs{1}{3}}{\eva}
\end{equation}
for the susceptibility, where \eq{eq:zi_plane_wave_form} was used for $\mb{E}$, $\mc{E}$ was substituted in terms of $\eva$, and $d$ was substituted in terms of $g$; $N$ is the ensemble density, and $N_{atoms}$ is the total number of atoms in the medium.

Because $g$ is in units of frequency, it will later be written in terms of some rate $\Gamma$ for the sake of convenience in the numerical calculations. Therefore, $\chi$ effectively depends on the factor $\Gamma/ck_0$, which is also present in other equations throughout this paper. We can use it as a scale factor that relates a dimensionless spatial variable with a dimensionless time variable, which then determines the speed of the pulse in the numerical simulation, and choose a value for it that makes the numerics more practical. This is why $\Gamma/ck_0$ will be specified in the following results.

The index is
\begin{equation}\label{eq:zi_three-level_index}
    \begin{split}
        n=\sqrt{\epsilon\mu}&=\sqrt{1+\chi} \\
        &=\sqrt{1+\frac{2N_{atoms}g}{ck_0}\frac{\evs{1}{3}}{\eva}} ,
    \end{split}
\end{equation}
where the relative permittivity is $1+\chi$, and the relative permeability is assumed to be $1$ because there is negligible magnetization for visible wavelengths in atomic media. The sign of $n$ must be initially chosen so that $\text{Im}(n)\ge0$, meaning that there is no gain, in the case of $\Gamma_{inc}=0$ to be physical. The goal is to obtain a susceptibility $\chi=-1$ through choice of parameters, so that the index is exactly equal to zero.

Next, Bloch equations are derived from the Lindblad master equation, $\dot{\hat{\rho}} = -\frac{i}{\hbar}\comm{\hat{H}}{\hat{\rho}} + \displaystyle\sum_{m}\Gamma_m\left[ \hat{L}_m \hat{\rho} \hat{L}_{m}^{\dagger} - \frac{1}{2} \left( \hat{L}_{m}^{\dagger}\hat{L}_{m}\hat{\rho} + \hat{\rho} \hat{L}_{m}^{\dagger} \hat{L}_{m} \right)\right]$ with the Hamiltonian \eq{eq:zi_three-level_hamiltonian}; the summation is over all the involved operators $\hat{L}_m$ ($\hat{a}$ and $\hat{\sigma}_{ij}$), with associated decay rates $\Gamma_m$. These are:
\begin{equation}\label{eq:zi_three-level_bloch}
    \begin{split}
        \pdv{\evs{1}{2}}{t} \approx& \;\left(i\Delta_1-i\Delta_2-\Gamma_{inc}/2\right)\evs{1}{2} - ig\eva\evs{2}{3}^* \\ 
        &+ i\Omega^*\evs{1}{3} , \\
        \pdv{\evs{1}{3}}{t} \approx& \;\left(i\Delta-\Gamma_1/2-\Gamma_2/2-\Gamma_{inc}/2\right)\evs{1}{3} \\ 
        &- ig\eva\left(\evs{3}{3}-\evs{1}{1}\right) + i\Omega\evs{1}{2} , \\
        \pdv{\evs{2}{3}}{t} \approx& \;\left(i\Delta_2-\Gamma_1/2-\Gamma_2/2\right)\evs{2}{3} + ig\eva\evs{1}{2}^* \\
        & - i\Omega\left(\evs{3}{3}-\evs{2}{2}\right) , \\
        \pdv{\evs{1}{1}}{t} \approx& \;ig\left(\evad\evs{1}{3} - \eva\evs{1}{3}^*\right) + \Gamma_1\evs{3}{3} \\ 
        &- \Gamma_{inc}\evs{1}{1} , \\
        \pdv{\evs{2}{2}}{t} \approx& \;i\Omega\left(\evs{2}{3} - \evs{2}{3}^*\right) + \Gamma_2\evs{3}{3} , \\
        \pdv{\evs{3}{3}}{t} \approx& \;ig\left(\eva\evs{1}{3}^* - \evad\evs{1}{3}\right) \\ 
        &+ i\Omega\left(\evs{2}{3}^* - \evs{2}{3}\right) - \left(\Gamma_1+\Gamma_2\right)\evs{3}{3} \\
        &+ \Gamma_{inc}\evs{1}{1} .
        \end{split}
\end{equation}
We make the mean-field approximation which assumes that there are no correlations between the field and atoms, and we neglect interactions among atoms, so that these equations contain only expectation values of single operators. These approximations are made in the first approach for the sake of simplicity in the numerical calculations to follow, partly to limit the number of equations that must be solved, and also because these calculations would otherwise involve some difficulty in determining appropriate boundary conditions for certain variables. This is described further in Sections \ref{sec:zi_numerical_simulation} and \ref{sec:zi_numerical_simulation_five-level}.
    
\begin{figure}
    \centering
    \includegraphics[trim={0cm 0cm 0cm 0cm},clip,width=\linewidth]{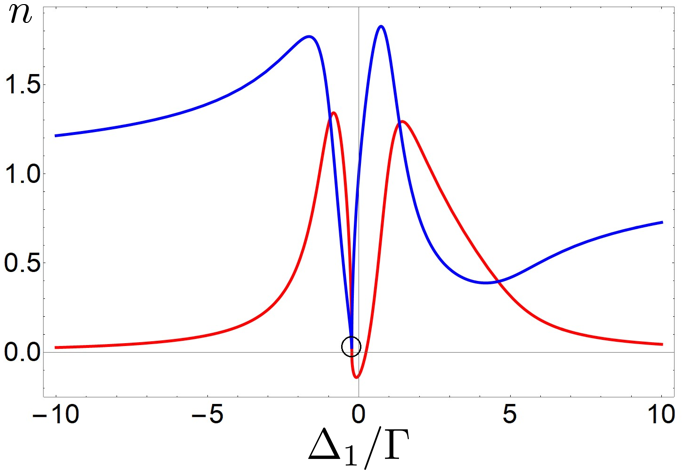}
	\caption{Refractive index as a function of $\Delta_1/\Gamma$. We chose the value for $\Delta_1$ where $\text{Im}(\evs{1}{3})=0$ and $\text{Re}(\evs{1}{3})$ is negative, which was $\Delta_1=-0.235\Gamma$. Then we chose $N_{atoms}$ such that $\chi=-1$ and therefore $n$ is near zero: $n=0.000552-3.103\times10^{-13}\text{i}$. The other parameters are: $\Delta_2=0\Gamma$, $g=0.01\Gamma$, $\Omega=1\Gamma$, $\Gamma_1=\Gamma_2=\Gamma$, and $\Gamma/ck_0=0.1$.}
     \label{fig:zi_three-level_index_plot}
\end{figure}

In order to find parameters that give $\chi=-1$, we solve \eqp{eq:zi_three-level_bloch} in the steady state. To do this, we assume that there is a constant effective Rabi frequency from the probe field throughout the medium, $\Omega_{eff}=g\eva(0,0)$, using the initial value of the pulse given by \eq{eq:zi_pulse} at the boundary. This is approximately as if the relatively flat tail of the pulse begins inside the medium from $z=0$ to $z=z_f$. We choose a set of typical parameters except for $N_{atoms}$ and $\Delta_1$, and solve for $\evs{1}{3}$ as if it were for a single atom. Results are shown in \fig{fig:zi_three-level_index_plot}.

The curve for $\text{Re}(\evs{1}{3})$ is similar to the well-known result for three-level atoms \cite{scully2008,loudon2000} without incoherent pumping, but the addition of incoherent pumping causes the dip in $\text{Im}(\evs{1}{3})$ below zero, centered at $\Delta_1=0$. We choose the point where $\text{Im}(\evs{1}{3})=0$ and $\text{Re}(\evs{1}{3})$ is negative as our working point. Here, there is no attenuation or gain, but the negative real part can be scaled to $-1$ by increasing the total number of atoms in the medium. Using \eq{eq:zi_three-level_susc}, we find that $N_{atoms}=594155$ gives $\chi=-1$ and therefore $n=0$.

The final necessary piece of the model is an equation for the propagation of $\eva$. First, we use Maxwell's equations $\nabla \cross \mb{E} = -\pdv{\mb{B}}{t}$ and $\nabla \cross \mb{H} = \pdv{\mb{D}}{t}$ in a medium with linear polarization $\mb{P}=\epsilon_0\chi\mb{E}$, together with the form \eq{eq:zi_plane_wave_form} for electric field and analogously for the polarization, to derive an equation for the evolution of $\mc{E}\propto\eva$ in the medium. However, because $n$ is a function of $z$ and $t$, this equation contains terms involving derivatives of $n$ and is practically extremely difficult to solve, even numerically. Although we are interested in treating the refractive index as a function of space and time so that it can affect the field in ``real time" and vice versa, at this stage we treat $n$ as a constant so that we can neglect any term involving a derivative of $n$. This would not affect the results if $n$ were to remain at its initial, zero value, or at least if it varied much more slowly than $\mc{E}$ or $\mc{P}$. If the results show that $n$ stays near zero, then this would suggest that this approximation is valid. If $n$ varies quickly in time or space away from zero, then this would suggest that this approximation is not valid, but this would confirm that the ZI response is fragile anyway. In any case, the finite-element method that we will use, described in Section \ref{sec:zi_numerical_simulation}, further justifies this approximation, because we will simulate the propagation over small steps in space and time where the index can be assumed to be constant. With this simplification, we obtain

\begin{equation}\label{eq:zi_three-level_maxwell_with_n}
    \begin{split}
        \pdv[2]{\mc{E}}{z} + 2ink_0 & \pdv{\mc{E}}{z} -n^2k_0^2\mc{E} - \epsilon_0\mu_0 \left( \pdv[2]{\mc{E}}{t} - 2i\nu\pdv{\mc{E}}{t} - \nu^2\mathcal{E} \right) \\
        & = \mu_0 \left( \pdv[2]{\mc{P}}{t} - 2i\nu\pdv{\mc{P}}{t} - \nu^2\mc{P} \right) .
    \end{split}
\end{equation}

At this point, the slowly varying envelope approximation (SVEA) is often made in order to drop certain terms when $nk_0\abs{\mc{E}} \ll \abs{\pdv{\mc{E}}{z}}$ or $\nu\abs{\mc{E}} \ll \abs{\pdv{\mc{E}}{t}}$ is valid. However, since here the index is supposed to be near zero, we cannot make the SVEA to neglect any spatial derivatives, although we use it to drop the two terms involving second-order time derivatives to get
\begin{equation}\label{eq:zi_three-level_maxwell_with_n_svea}
    \begin{split}
        \pdv[2]{\mc{E}}{z} + 2ink_0 & \pdv{\mc{E}}{z} -n^2k_0^2\mc{E} + \epsilon_0\mu_0 \left( 2i\nu\pdv{\mc{E}}{t} + \nu^2\mathcal{E} \right) \\
        & = \mu_0 \left( - 2i\nu\pdv{\mc{P}}{t} - \nu^2\mc{P} \right) .
    \end{split}
\end{equation}
Finally, $\Emcal$ is replaced with $\sqrt{2\hbar\nu/\epsilon_0V}\eva$, $\mc{P}$ is replaced with $2Nd\evs{1}{3}$, and $d$ is put in terms of $g$ to arrive at
\begin{equation}\label{eq:zi_three-level_maxwell_a}
    \begin{split}
        \frac{1}{k_0^2} & \pdv[2]{\eva}{z}  + \frac{2in}{k_0}\pdv{\eva}{z} + \frac{2i}{ck_0}\pdv{\eva}{t} + \left( 1-n^2 \right) \eva \\
        &= - \frac{4iN_{atoms}g}{c^2k_0^2}\pdv{\evs{1}{3}}{t} - \frac{2N_{atoms}g}{ck_0}\evs{1}{3} .
    \end{split}
\end{equation}
The system of \eqp{eq:zi_three-level_bloch} and \eq{eq:zi_three-level_maxwell_a} must be solved numerically, with the parameters and initial conditions required for $\chi=-1$.

%
%
%
%
%
%
%
%
\subsection{Numerical Simulation Description}\label{sec:zi_numerical_simulation}
%
%
%
The set of Maxwell-Bloch equations \eqp{eq:zi_three-level_bloch} and \eq{eq:zi_three-level_maxwell_a} must be solved numerically in space and time with appropriate boundary conditions in $z$ and initial conditions in $t$. Some of these conditions are known at $z=0$ and at $t=0$, but none at other points in $z$ or $t$, where we cannot place any constraints on the system. For $\eva$, we have boundary conditions at $z=0$ and initial conditions at $t=0$, supplied by \eq{eq:zi_pulse} as the pulse enters the medium. For the atomic variables $\evs{i}{j}$, initial conditions can be found by solving the steady-state Bloch equations before the pulse enters, but we do not have any boundary conditions. Also, we want to treat $n$ as a variable during the propagation in order to allow it to be affected by $\eva$ and vice versa. However, the Maxwell equations were derived assuming that $n$ is constant, so we must simulate the system in a way that is consistent with this derivation while also allowing the index to vary.

\begin{figure}
    \centering
    \includegraphics[trim={0cm 0cm 0cm 0cm},clip,width=\linewidth]{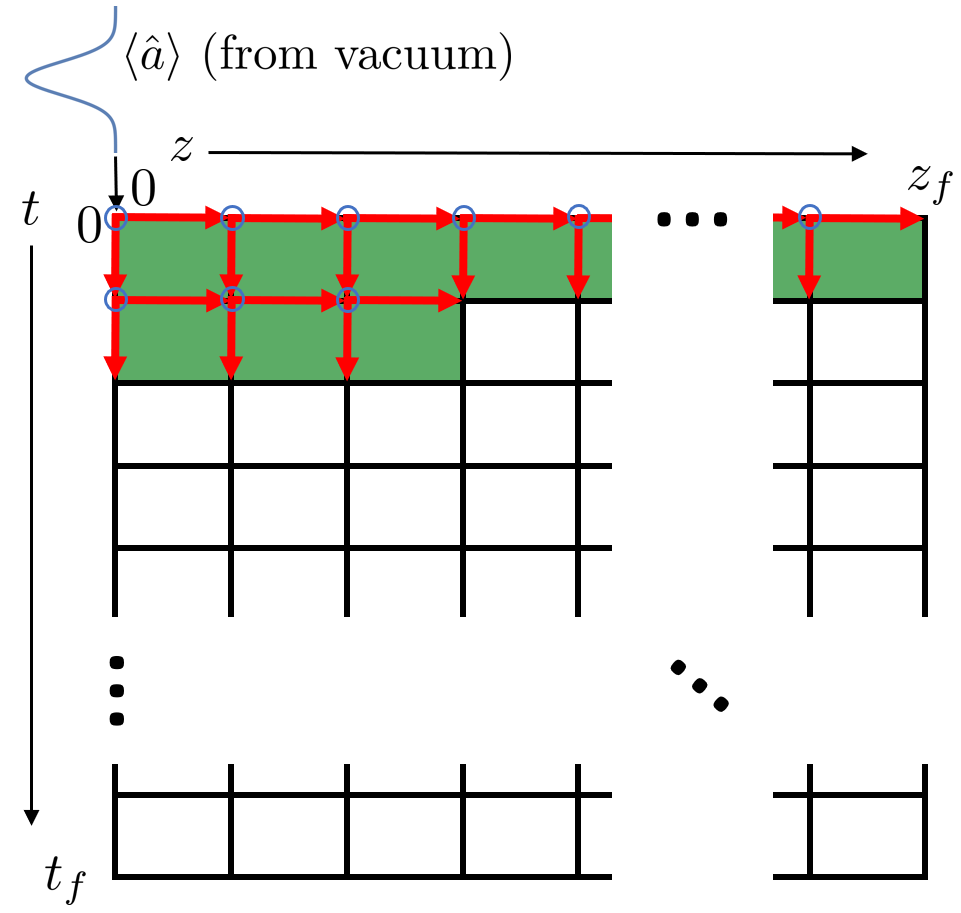}
	\caption{Illustration of numerical procedure. The $z-t$ region is divided into rectangles of side length $dz$ and $dt$. Each corner of each box lies at a point on a grid. At $z=0$, boundary values are taken from the pulse of \eq{eq:zi_pulse}. At $t=0$, initial values are obtained from the pulse for $\eva$, or steady-state solutions to the Bloch equations \eq{eq:zi_three-level_bloch} or \eq{eq:zi_five-level_bloch}. Otherwise, boundary conditions are obtained from previous solutions. The program manually steps along $z$, then $t$. At each blue circle, NDSolve is initialized and run from that $(z,t)$ coordinate, out to $z+dz$ and $t+dt$ to solve \eq{eq:zi_three-level_bloch} and \eq{eq:zi_three-level_maxwell_a}, or \eq{eq:zi_five-level_bloch} and \eq{eq:zi_five-level_maxwell_a} from Section \ref{sec:zi_five-level_model}, until solutions are obtained over the entire region. The filled-in green boxes indicate regions where NDSolve has given a solution.}
     \label{fig:zi_grid_numerics}
\end{figure}

These complications on top of an already complicated system of equations make it very difficult to solve. To get past these issues, we use a custom finite-element algorithm in conjunction with NDSolve in Mathematica. This procedure is illustrated in \fig{fig:zi_grid_numerics}. We imagine a grid over the $z-t$ region where the system is to be solved, with boxes of size $dz \times dt$. NDSolve is used to evolve the system over the region of each box individually, stepping along $z$ and then $t$; beginning at one corner of a box, NDSolve gives a solution out to $dz$ and $dt$. Boundary and initial conditions are updated at each step. At $z=0$, we use \eq{eq:zi_pulse} for boundary conditions for $\eva$. The pulse is placed far enough outside the medium so that the value of the ``tail" at $z=0$, $t=0$ is small; in all cases described in the results of Sections \ref{sec:zi_three-level_results} and \ref{sec:zi_five-level_results}, this value is about $0.003$. Inside the medium, we use values from solutions at previous steps for boundary conditions. 
At the boundary $z=0$, we have a continuous function for $\eva$. To reduce numerical error and allow the atomic variables to evolve realistically under the effect of $\eva$ at $z=0$, we use the Bloch equations to evolve them with the continuous pulse function, instead of using constant values for $\eva$ supplied step-by-step.

The instantaneous index $n$ is calculated after each step according to \eq{eq:zi_three-level_index}, with current values for the field and atomic variables. The current value for $n$ at that $z,t$ point on the grid is replaced as a constant in the Maxwell equation for the NDSolve calculation that originates at that point. For small step sizes, we can justify approximating $n$ as constant, which in turn justifies the approximations made by neglecting derivatives of $n$ in the derivation of the Maxwell equation \eq{eq:zi_three-level_maxwell_a}.

In numerical calculations, we use dimensionless variables $\zeta=k_0 z$ and $\tau=\Gamma t$ and rewrite the Maxwell-Bloch equations in terms of these variables. This is how the scale factor $\Gamma/ck_0$ appears and why it is specified for the results shown in the rest of the paper.

We set the spatial step size $d\zeta=k_0dz$ by choosing the number of steps in space, zSteps, and we set the temporal step size $d\tau=\Gamma dt$ by choosing the number of steps in time, tSteps. The relative step size $d\zeta/d\tau$ effectively sets the speed of propagation in the medium in the simulation.

\subsection{Results}\label{sec:zi_three-level_results}
We now present results for numerical simulations of the model described in Section \ref{sec:zi_three-level_model} via the method described in Section \ref{sec:zi_numerical_simulation}.

Results are given below for the same parameters as in \fig{fig:zi_three-level_index_plot}, with additional plots in Appendix \ref{3levelappendix}. They show that the instantaneous refractive index $n$ given by \eq{eq:zi_three-level_index} immediately moves away from its initial near-zero value. In the first time step, at $z=0$, the real part jumps to about $0.66$ and the imaginary part jumps to about $0.3$. The index then continues to vary for low $z$, just inside the medium, as long as part of the pulse is entering. The real part of $n$ goes from a maximum of just over $1.0$ down to a minimum of about $-2.3$ at $z=0$. Further inside the medium, the extent of this variation generally decreases until the index settles near zero again, even where there is a significant amount of the pulse present.

\begin{figure}
    \centering
    \includegraphics[trim={0cm 0cm 0cm 0cm},clip,width=\linewidth]{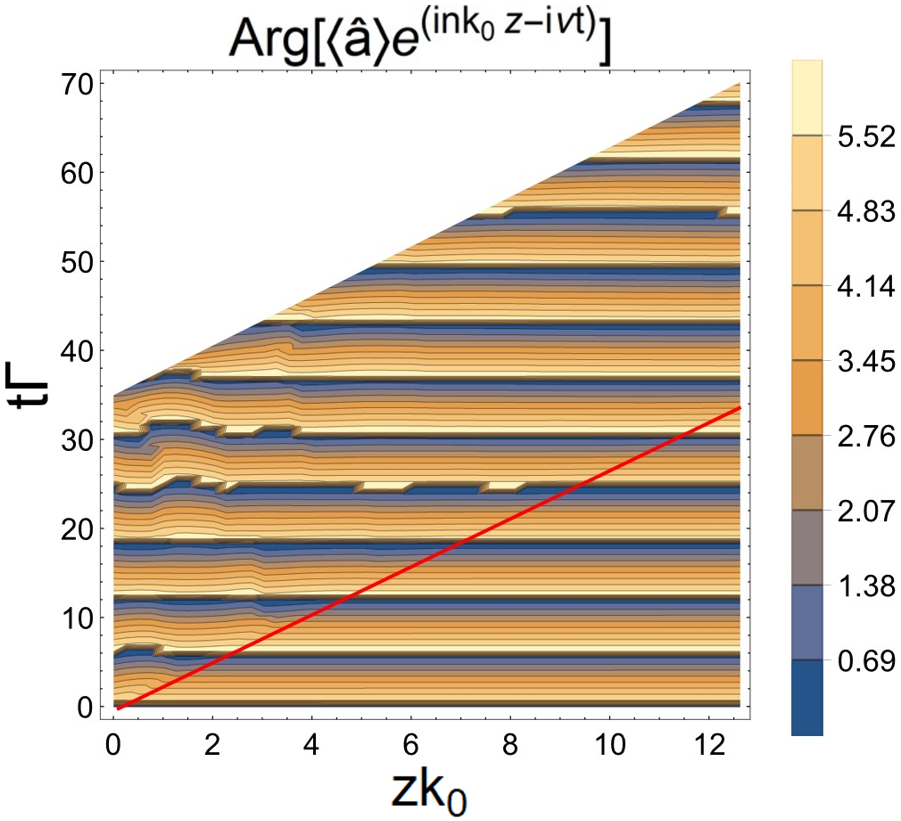}
	\caption{Phase of $\eva e^{ink_0z-i\nu t}$. The index does not remain near zero. Along $t$, there is regular phase variation, but there is also spatial phase change due to the nonzero index. This is especially noticeable below $zk_0=6$. Above that, the phase does not significantly change in $z$. The red line indicates the path of the beginning of the pulse; below the line, the system is in the prepared ZI steady state. The white region in the upper left corner is where the simulation was cut off. Parameters are: $N_{photons}=1$, $\sigma_s=10\pi/k_0$, $z_i=-120/k_0$, $\Gamma_1=\Gamma_2=\Gamma$, $\Gamma_{inc}=0.3\Gamma$, $\Delta_1=-0.235\Gamma$, $\Delta_2=0$, $g=0.01\Gamma$, $\Omega=\Gamma$, $\Gamma/(ck_0)=0.1$, and $N_{atoms}=594155$. We set $z_f=4\pi/k_0$, zSteps=50, $t_f=70/\Gamma$, and tSteps=100.}
     \label{fig:3La_c_arg}
\end{figure}

The phase of $\eva e^{ink_0z-i\nu t}$ is shown in \fig{fig:3La_c_arg}. There is some spatial variation for low $z$ where the index changes, especially below $zk_0=6$, and there is a slight change along the line $t\Gamma = (70/8\pi)zk_0$ (just above the red line in \fig{fig:3La_c_arg}), which is the path of the front ``tail" of the pulse that enters the medium at $z=0$, $t=0$. The change along this line is due to a small imaginary part that $\eva$ gains after entering the medium, shown in \fig{fig:3La_c_a}. For most of the $z-t$ region, there is little spatial variation because the index (both instantaneous and effective) is near zero, and the field envelope $\eva$ travels through the medium with very little change. The full field is shown in \fig{fig:3La_c_fullfield}, where these effects are also seen.

The coherence of the transition coupled to the probe field, $\evs{1}{3}$, roughly follows the envelope $\eva$, but near the boundary $z=0$, there is a delay which is the cause of the index variation, which can be seen from \eq{eq:zi_three-level_susc}.

\begin{figure}
    \centering
     \label{fig:3La_c_fullfield}
\end{figure}

\begin{figure}
    	\includegraphics[trim={0cm 0cm 0cm 0cm},clip,width=\linewidth]{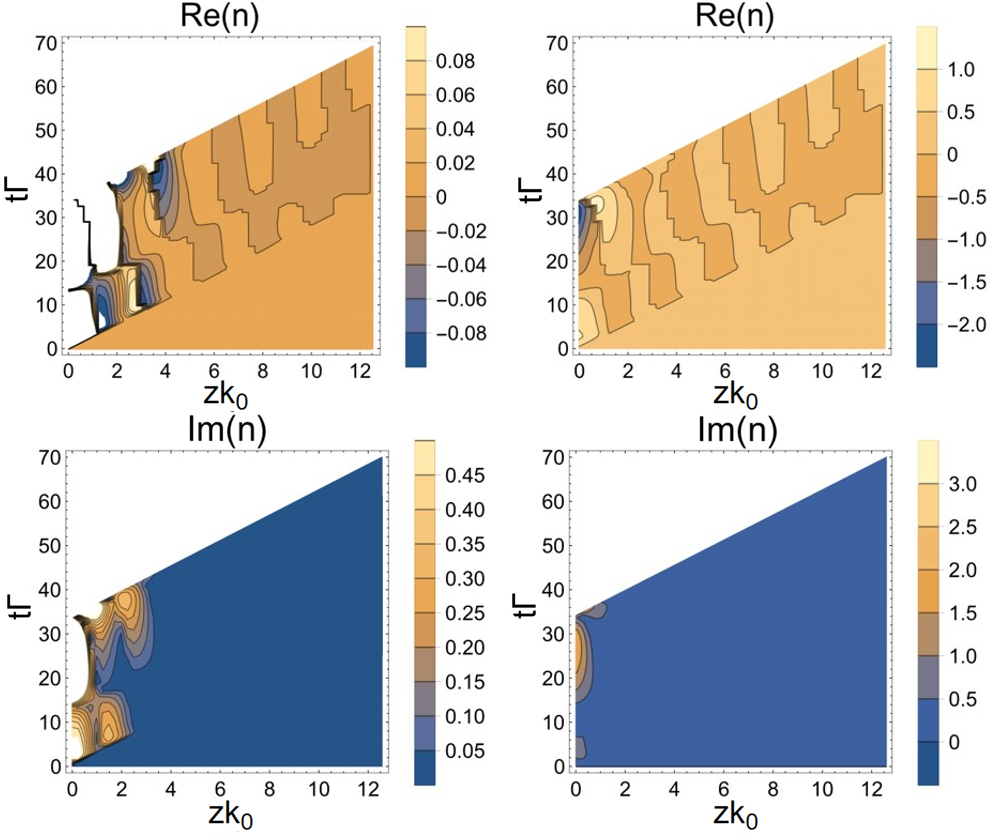}
	\caption{$n$ in time and space, with three different scales each for the real and imaginary parts. Both the real and imaginary parts change from the initial zero value most drastically near $z=0$, before settling towards zero for larger $z$. The white regions in the upper left corners are where the simulation was cut off. Parameters are the same as in \fig{fig:3La_c_arg}.}
\end{figure}

\subsection{Discussion}\label{sec:zi_three-level_discussion}
The refractive index does not remain zero as a pulse of light propagates into the medium consisting of three-level atoms, even for a single photon. The index rises from zero in the first time step. As $\eva$ grows at the boundary to the medium, $\evs{1}{3}$ does not evolve at the same rate; even though the atomic variables evolve continuously according to the Bloch equations \eq{eq:zi_three-level_bloch} with $\eva$ given by \eq{eq:zi_pulse} for a pulse, their evolution is slower, so the balance is broken, and the susceptibility $\chi$ immediately changes from $-1$. Further into the medium, the index stays nearly zero as a single-photon pulse travels through it. It is interesting that this happens not later in time near the boundary as the coherence builds up there, but further in space, even where the pulse is just beginning to reach.

Other simulations showed that the variation in index is slightly reduced for a wider pulse, which has a slower rate of change at the boundary, and is slightly increased for pulses containing more than one photon. Also, if the incoming field rises to some steady value, the index varies while the envelope varies, but after the envelope reaches a steady value, the index does as well, which can be near zero for a small field amplitude or a slow increase.

The variation in index is not particularly surprising, since a zero instantaneous index as given by \eq{eq:zi_three-level_index} relies on $\eva$ and $\evs{1}{3}$ to remain in the same proportion, as precisely prepared via choice of $\Delta_1$, $N_{atoms}$, and other parameters. By the Kramers-Kronig relations, the refractive index varies quickly with frequency; therefore, after being prepared near zero for a single frequency, it likely quickly moves away from zero if the light affects the medium's response at all. This can be seen from \fig{fig:zi_three-level_index_plot}, where the susceptibility varies quickly with frequency at the working point chosen; the slope of $\text{Re}(\chi)$ is large at the chosen value of $\Delta_1$.

These observations suggest that any field with a time-dependent amplitude near the boundary ruins its ZI, even for a slowly varying amplitude, and with as little light as one photon. Even though the instantaneous index may stay near zero in part of the medium, the field will experience some phase change in space during its propagation. The phase plots show a slight additional phase change due to the effective index, represented by a small imaginary part gained by $\eva$. However, if the amplitude settles at a constant value in the medium, then the coherence and index will as well, and the effective index can still be near zero through the entire medium. For a larger maximum field amplitude, or rather a faster increase to this maximum, the index settles further from zero.

%
%
%
%
\section{Propagation Through Five-Level Atoms}

\subsection{Theoretical Model}\label{sec:zi_five-level_model}
\begin{figure}
    \centering
    \includegraphics[trim={0cm 0cm 0cm 0cm},clip,width=\linewidth]{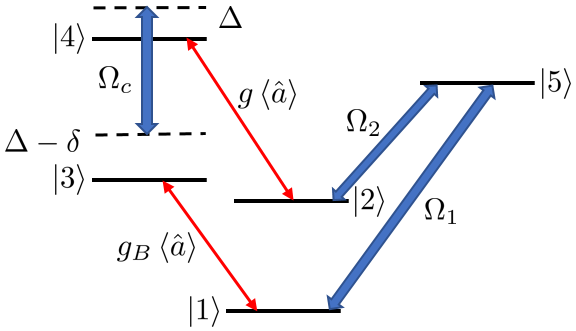}
	\caption{Five-level atom coupled to external fields. $g\eva$ and $g_B\eva$ are effective Rabi frequencies on transitions coupled to the probe field. $\Omega_1$, $\Omega_2$, and $\Omega_3$ are pump-field Rabi frequencies. $\Delta-\delta$ and $\Delta$ are detunings.}
     \label{fig:zi_five-level_atom}
\end{figure}

Since the index is apparently fragile for the three-level atomic medium due to the strict requirement placed on $\chi$ and its sensitivity to the probe frequency, we move on to a medium consisting of the five-level atoms shown in \fig{fig:zi_five-level_atom}. This structure was chosen because previous work has shown that it is capable of giving a negative-index response to a probe field with low attenuation, so by scaling the density or varying other parameters, a ZI response should also be possible \cite{kastel2007,orth2013}. In this case no incoherent pumping is required to achieve ZI like in the three-level structure presented in Section \ref{sec:zi_three-level_model}. Also, the pump fields largely set the response to the probe field, which has less of an effect on the atomic variables than for the three-level structure. This suggests that it may be easier to maintain the ZI.

An electric field $\mb{E}$ and a magnetic field $\mb{B}$, belonging to the same ``probe field," each couple to a transition: the magnetic field couples to the $\trn{1}{3}$ transition which has the magnetic dipole operator $\hat{\mb{m}}$ (assumed real) and coupling strength $g_B$, and the electric field couples to the $\trn{2}{4}$ transition which has the electric dipole operator $\hat{\mb{d}}$ (assumed real) and coupling strength $g$. These transitions have the same characteristic frequency. There are pump fields with Rabi frequencies $\Omega_1$, $\Omega_2$, and $\Omega_3$ on the other three transitions.

The detunings are $\Delta=\nu-\left( \omega_4-\omega_1 \right)$ and $\delta=\nu_{\Omega_3}-\left( \omega_4-\omega_3 \right)$, where $\nu_{\Omega_3}$ is the frequency of the pump field on $\trn{3}{4}$, and $\omega_i$ is the frequency corresponding to the atomic state $\ket{i}$. The decay rates involved are $\Gamma_{31}$, $\Gamma_{42}$, $\Gamma_{43}$, $\Gamma_{51}$, and $\Gamma_{52}$, where $\Gamma_{ij}$ is the decay rate from $\ket{i}$ to $\ket{j}$. The single-atom Hamiltonian is
\begin{equation}\label{eq:zi_five-level_hamiltonian}
    \begin{split}
        \hat{H}/\hbar &= (\delta-\Delta)\hat{\sigma}_{33} - \Delta\hat{\sigma}_{44} - ( g\hat{a}\hat{\sigma}_{42} \\
        &+ g_B\hat{a}\hat{\sigma}_{31} + \Omega_1\hat{\sigma}_{51} + \Omega_2\hat{\sigma}_{52} + \Omega_3\hat{\sigma}_{42} + \text{H.c.} ) .
    \end{split}
\end{equation}

Without any detuning in the $\ket{1}$-$\ket{2}$-$\ket{5}$ subsystem, there is two-photon resonance, and electromagnetically induced transparency occurs which creates a dark superposition of states $\ket{1}$ and $\ket{2}$ \cite{scully2008,kastel2007}. Essentially all of the population accumulates in this dark state. For a probe field on the two transitions from the dark state, the response can be treated to linear order in $\mb{E}$ and $\mb{B}$ to a good approximation, as the probe field only perturbs the atom from the dark state.

The field $\Omega_3$ induces a cross-coupling between electric and magnetic components of the probe field which is key to a zero (or negative) index. This cross-coupling is seen in the form of the induced polarization and the induced magnetization, which are most generally

\begin{equation}\label{eq:zi_p_m_tensor}
    \begin{split}
        \mb{P}=\epsilon_0\tens{\chi}_{ee}\mb{E}+c\epsilon_0\tens{\chi}_{eb}\mb{B}, \\
        \mb{M}=c\epsilon_0\tens{\chi}_{be}\mb{E}+\frac{\tens{\chi}_{bb}}{\mu_0}\mb{B},
    \end{split}
\end{equation}
where the $\tens{\chi}_{ij}$ are susceptibility tensors. This means that the calculation of the index is not as straightforward as in Section \ref{sec:zi_three-level_model}. Also, there will be additional terms in the Maxwell equation due to the presence of magnetization, and the Bloch equations will be more complicated due to the additional levels.

First, a new expression for the refractive index must be found. Using $\curl{\mb{E}}=-\pdv{\mb{B}}{t}$ and $\curl{\mb{H}}=\pdv{\mb{D}}{t}$, all but $\mb{E}$ are eliminated. Assuming the form \eq{eq:zi_plane_wave_form} with a completely general polarization, the tensor equation is reduced to a scalar equation which can be solved for $n$, but two approximations are made. First, $n$ is treated as a constant for the reasons described in Section \ref{sec:zi_three-level_model}. Second, the amplitude $\mc{E}$ is assumed to be constant; otherwise the expression for $n$ would contain many terms with derivatives of $\mc{E}$. This approximation is not necessary, but made here simply to make the following numeric simulations more tractable; however, this is justified by the numerical method for the same reason that treating $n$ as a constant is justified. The result is
\begin{equation}\label{eq:zi_five-level_index}
    n=\frac{\left( \chi_{eb}+\chi_{be}\right) \pm \sqrt{4\left( 1+\chi_{ee}\right)\left( 1-\chi_{bb}\right) + \left( \chi_{eb}+\chi_{be}\right)^2}}{2\left( 1-\chi_{bb}\right)} ,
\end{equation}
where, for example, $\chi_{eb}=\hat{\epsilon}_E^\dagger \tens{\chi}_{eb} \hat{\epsilon}_B$. The four $\chi_{ij}$ must be calculated using the Bloch equations, and a set of parameters found such that $\text{Re}(n)$ is near zero.

\begin{figure}
    \centering
    \includegraphics[trim={0cm 0cm 0cm 0cm},clip,width=\linewidth]{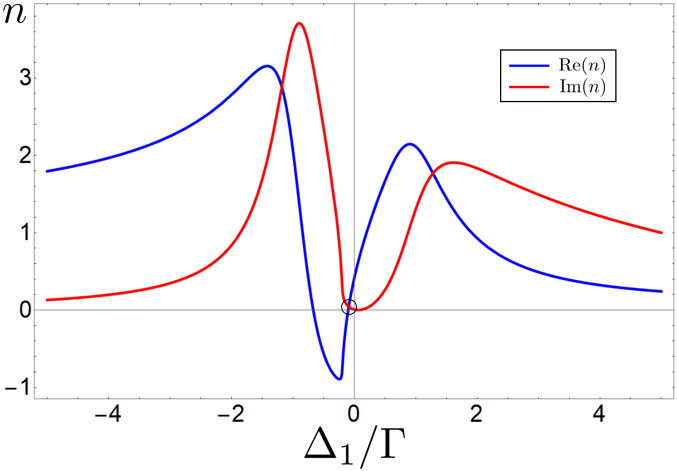}
	\caption{Refractive index as a function of $\Delta/\Gamma$. At $\Delta=-0.0993\Gamma$, $\text{Re}(n)=0$ and $\text{Im}(n)=0.0498$ (circled). Other parameters are $N_{atoms}=5\times10^5$, $\delta=0$, $g=0.01\Gamma$, $g_B=0.001\Gamma$, $\Omega_1=\Omega_2=\Omega_3=\Gamma$, $\Gamma_{31}=(1/137)^2\Gamma$, $\Gamma_{42}=\Gamma_{43}=\Gamma_{51}=\Gamma_{52}=\Gamma$, and $\Gamma/ck_0=0.1$.}
     \label{fig:zi_five-level_index_plot}
\end{figure}

\eqp{eq:zi_p_m_tensor} can also be reduced to scalar form:
\begin{equation}\label{eq:zi_p_m_scalar}
    \begin{split}
        \mc{P}=\epsilon_0\chi_{ee}\mc{E} + c\epsilon_0\chi_{eb}\mc{B} = 2Nd\evs{2}{4}, \\
        \mc{M}= c\epsilon_0\chi_{be}\mc{B} + \frac{\chi_{bb}}{\mu_0}\mc{B} = 2Nm\evs{1}{3}.
    \end{split}
\end{equation}
The coherences $\evs{1}{3}$ and $\evs{2}{4}$ are solved for in the steady state in terms of other $\evs{i}{j}$, to first order in the probe field. Then the four $\chi_{ij}$ can be identified. In terms of $g$ and $g_B$ rather than $d$ and $m$, they are
\begin{equation}\label{eq:zi_five-level_chis}
    \begin{split}
        \chi_{ee} & = \frac{8N_{atoms} g^2}{ck_0} \\
        &\times \frac{i\left[ \Gamma_{31}+2i(\delta-\Delta)\right] \evs{2}{2}}{\left[ \Gamma_{31}+2i(\delta-\Delta)\right]\left(\Gamma_{42}+\Gamma_{43}-2i\Delta\right) + 4\abs{\Omega_3}^2}, \\
        \chi_{eb} & = \frac{16N_{atoms} g g_B}{ck_0} \\
        &\times \frac{\Omega_3 \evs{2}{1}}{\left[ \Gamma_{31}+2i(\delta-\Delta)\right]\left(\Gamma_{42}+\Gamma_{43}-2i\Delta\right) + 4\abs{\Omega_3}^2}, \\
        \chi_{be} & = \frac{16N_{atoms} g g_B}{ck_0} \\
        &\times \frac{\Omega_3^* \evs{1}{2}}{\left[ \Gamma_{31}+2i(\delta-\Delta)\right]\left(\Gamma_{42}+\Gamma_{43}-2i\Delta\right) + 4\abs{\Omega_3}^2}, \\
        \chi_{bb} & = \frac{8N_{atoms} g_B^2}{ck_0} \\
        &\times \frac{i\left(\Gamma_{42}+\Gamma_{43}-2i\Delta\right) \evs{1}{1}}{\left[ \Gamma_{31}+2i(\delta-\Delta)\right]\left(\Gamma_{42}+\Gamma_{43}-2i\Delta\right) + 4\abs{\Omega_3}^2}.
    \end{split}
\end{equation}
\eq{eq:zi_five-level_index} and \eq{eq:zi_five-level_chis} suggest that the index experienced by the probe field will be fairly stable, since the response is mostly determined by $\evs{1}{1}$, $\evs{2}{2}$, and $\evs{1}{2}^{(*)}$, which are largely fixed by the pump fields, even without the presence of the probe field. \eqp{eq:zi_five-level_chis} are substituted into \eq{eq:zi_five-level_index} and parameters are chosen to obtain a ZI. An example is shown in \fig{fig:zi_five-level_index_plot}. We get $\text{Re}(n)=0$ at $\Delta=-0.0993\Gamma$, and also $\text{Im}(n)=0.0498$, which means there is fairly low attenuation; in any case, attenuation or gain will affect the amplitude of the field, but not its phase, so this does not matter for the present purpose.

Next, Bloch equations are derived from the Lindblad master equation with \eq{eq:zi_five-level_hamiltonian}, with the mean-field approximation as in Section \ref{sec:zi_three-level_model}. These are given in Appendix \ref{five-level_bloch}.

Finally, there is a contribution to the Maxwell equation due to the magnetization. After reducing the full Maxwell's equations to one scalar equation, we have
\begin{equation}\label{eq:zi_five-level_maxwell_scalar}
    \nabla^2 E - \epsilon_0\mu_0 \pdv[2]{t}E = \mu_0 \pdv[2]{t}P + \mu_0 \pdv{t}\pdv{z}M,
\end{equation}
where $E(z,t)=\frac{1}{2}\Emcal(z,t) e^{in(z,t)k_0z-i\nu t}$ and analogously for $P$, $M$, and $B$. These forms are substituted in \eq{eq:zi_five-level_maxwell_scalar}, then $\Emcal$ is replaced with $\sqrt{2\hbar\nu/\epsilon_0 V}\eva$, $\mc{P}$ and $\mc{M}$ with \eqp{eq:zi_p_m_scalar}, and $\chi_{ij}$ with \eqp{eq:zi_five-level_chis}. Finally, with the same approximations made as in Section \ref{sec:zi_three-level_model}, we obtain
\begin{equation}\label{eq:zi_five-level_maxwell_a}
    \begin{split}
        \frac{1}{k_0^2}&\pdv[2]{\eva}{z} + \frac{2in}{k_0}\pdv{\eva}{z} + \frac{2i}{ck_0}\pdv{\eva}{t} + \left( 1-n^2 \right) \eva \\
        &= - \frac{4iN_{atoms}g}{c^2k_0^2}\pdv{\evs{2}{4}}{t} - \frac{2N_{atoms}g}{ck_0}\evs{2}{4} \\
        +& \frac{2in N_{atoms} g_B}{(ck_0)^2}\pdv{\evs{1}{3}}{t} + \frac{2in N_{atoms} g_B}{ck_0}\evs{1}{3} \\
        -& \frac{2iN_{atoms} g_B}{ck_0}\pdv{\evs{1}{3}}{z} + \frac{2N_{atoms} g_B}{c^2k_0}\pdv{\evs{1}{3}}{z}{t}.
    \end{split}
\end{equation}

%
%
%
%

\subsection{Numerical Simulation}\label{sec:zi_numerical_simulation_five-level}
To solve the set of Maxwell-Bloch equations, \eqp{eq:zi_five-level_bloch} and \eq{eq:zi_five-level_maxwell_a}, we use the method described in Section \ref{sec:zi_numerical_simulation} for all the reasons described there. The instantaneous index $n$ is calculated after each step according to \eq{eq:zi_five-level_index}. 

However, there is an additional difficulty in this case. This comes from the terms involving $\evs{1}{3}$, related to the magnetization, in \eq{eq:zi_five-level_maxwell_a}. These include spatial derivatives of an atomic variable, which means a boundary condition in $z$ is required to solve the equations. For atomic variables, we only have valid initial conditions in $t$. Setting a boundary condition for an atomic variable, for example, at $z=0$, implies that we know the value of that variable at $z=0$ for all time, which is a constraint that we cannot justifiably place. 

Instead, we use the finite-difference method to approximate these terms at each step, using values from previous solutions. At $z=0$, $t=0$, we do not have previous solutions for $\evs{1}{3}$, so we approximate by solving the Bloch equations in the steady state using values of the pulse at $-dz$ and/or $-dt$. Furthermore, looking to future work, in order to include higher-order expectation values like $\expval{\hat{a}^{(\dagger)}\hat{\sigma}_{ij}}$ as a better approximation, this kind of approach would be necessary for dealing with spatial derivatives of these variables for any atomic medium.

\subsection{Results}\label{sec:zi_five-level_results}
We present results for the model described in Section \ref{sec:zi_five-level_model}, using the numerical method described in Sections \ref{sec:zi_numerical_simulation} and \ref{sec:zi_numerical_simulation_five-level}.

\begin{figure}
    \centering
    \includegraphics[trim={0cm 0cm 0cm 0cm},clip,width=\linewidth]{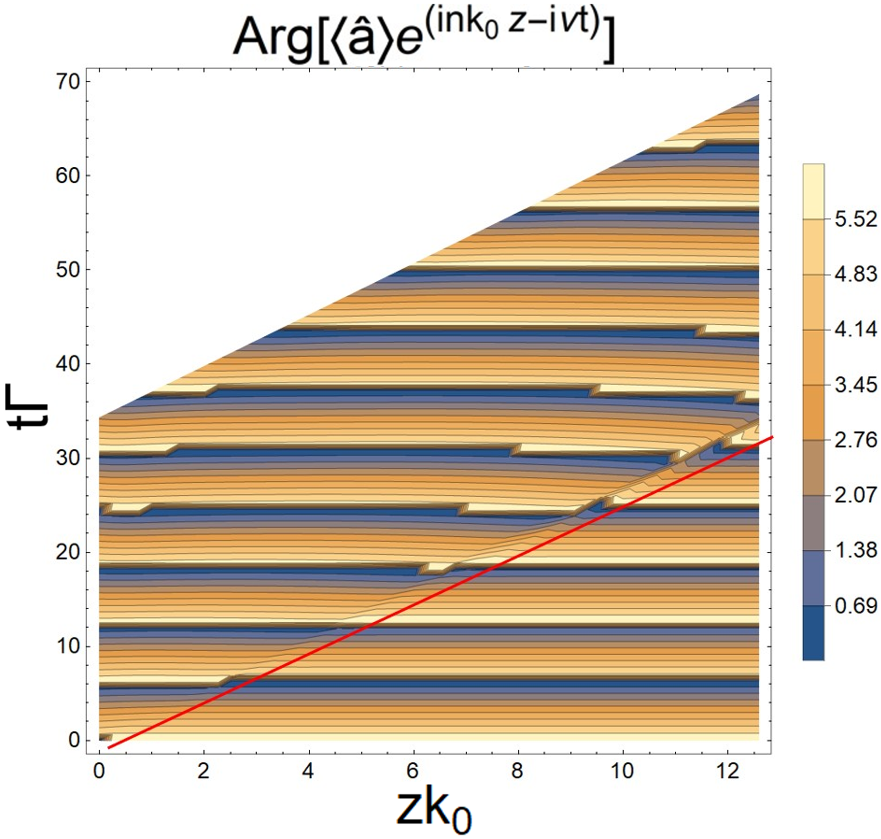}
	\caption{Phase of $\eva$. The index does not remain near zero. There is a spatial phase change right above the red line, along the path of the beginning of the pulse. This becomes sharper the further the pulse travels into the medium. This is caused by the effective index, which leads to a growing imaginary part of $\eva$. Besides that, there is little to no variation along $z$. Parameters are $N_{photons}=1$, $\sigma_s=10\pi/k_0$, $z_i=-120/k_0$, $\Gamma_{31}=\Gamma/137^2$, $\Gamma_{42}=\Gamma_{43}=\Gamma_{51}=\Gamma_{52}=\Gamma$, $\Delta=-0.0993\Gamma$, $\delta=0$, $g=0.01\Gamma$, $g_B=0.001\Gamma$, $\Omega_1=\Omega_2=\Omega_3=\Gamma$, and $N_{atoms}=5\times10^5$ (the same as in \fig{fig:zi_five-level_index_plot}). We set $z_f=4\pi/k_0$, zSteps=50, $t_f=70/\Gamma$, and tSteps=100.}
     \label{fig:5La_c_arg}
\end{figure}

The results show that the effective index does not remain zero, as there is spatial phase change in \fig{fig:5La_c_arg}, most noticeably along the path of the front tail of the pulse beginning at $z=0$, $t=0$, and $\eva$ grows even though $\text{Im}(n)>0$. However, compared to the three-level results, the instantaneous index stays much closer to zero overall, and the effective index stays closer to zero over a larger $z-t$ region.

\begin{figure}
    \centering
    \includegraphics[trim={0cm 0cm 0cm 0cm},clip,width=\linewidth]{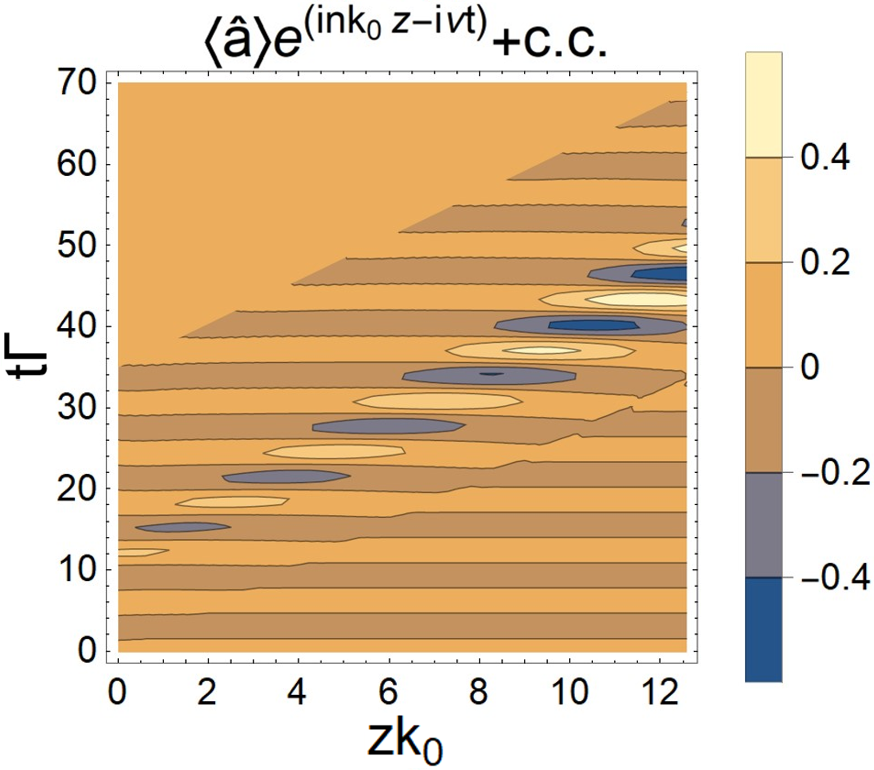}
	\caption{The full field in time and space (in arbitrary units). The pulse widens and its amplitude grows as it travels further into the medium. The effect of the overall index is seen along the path of the beginning of the pulse and just above it. Parameters are the same as in \fig{fig:5La_c_arg}.}
     \label{fig:5La_c_fullfield}
\end{figure}

\begin{figure}
\centering
    	\includegraphics[trim={0cm 0cm 0cm 0cm},clip,width=\linewidth]{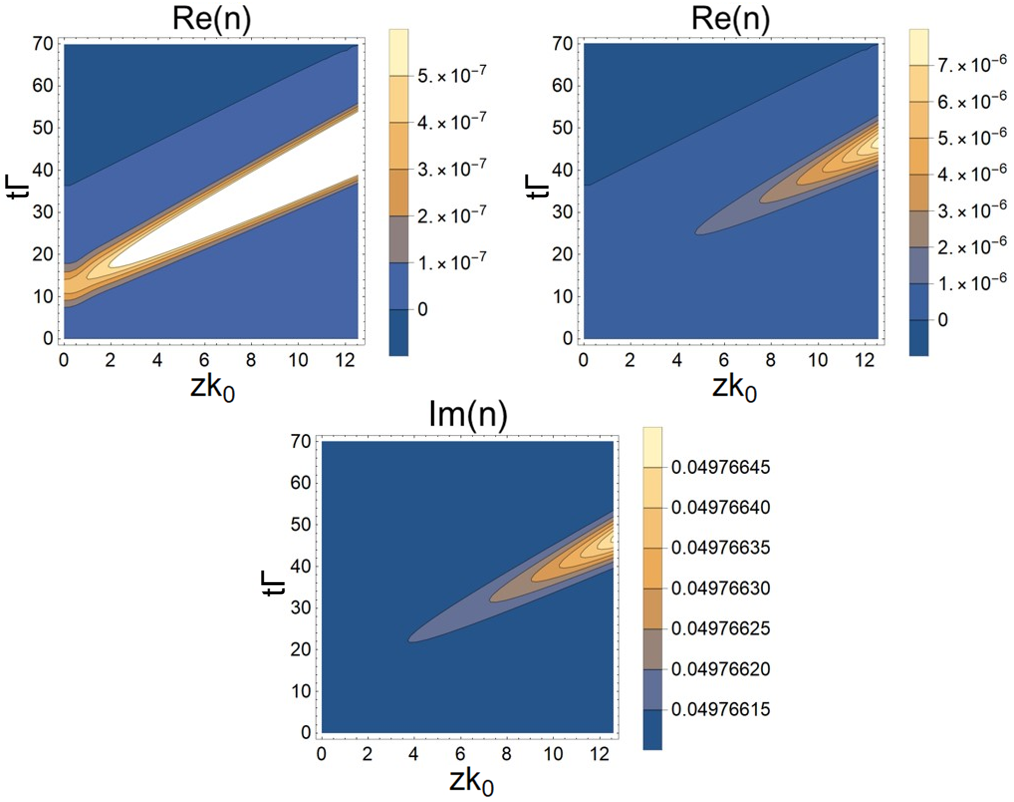}
	\caption{$n$ in time and space, with two different scales for the real part. Each part stays near its initial value, but changes slightly with the field as a growing $\eva$ perturbs the atomic medium further from its initial steady state. Parameters are the same as in \fig{fig:5La_c_arg}.}
     \label{fig:5La_c_n}
\end{figure}

\subsection{Discussion}\label{sec:zi_five-level_discussion}
There is a clear disturbance in the phase along the line $t\Gamma = (70/8\pi)zk_0$, where the initial part of the tail of the pulse propagates. This suggests that despite the near ZI preparation, the initial interaction between the photon and the medium causes a sudden phase change which is carried through the rest of the propagation. Mathematically, the spatial phase change in this case comes more from the imaginary part of $\eva$ that grows from the beginning of the propagation (see \fig{fig:5La_c_a}), rather than from the instantaneous $n$. This is also present in the three-level results of Section \ref{sec:zi_three-level_results}, but is less pronounced there since the imaginary part grows much more slowly. 

Simulations for other parameters suggest that this sharp phase change can be reduced, but not eliminated, with a wider pulse or by beginning with the peak of the pulse further outside the medium. There is a less prominent phase change above that line, at later times, when the pulse grows to its peak and then decreases. As in Section \ref{sec:zi_three-level_results}, the index likely varies from its initial value according to the Kramers-Kronig relations. The spatial phase variations suggest that even a single photon does ruin its effective ZI with the five-level atoms as well.

Despite this phase change, the instantaneous index, as calculated from \eq{eq:zi_five-level_index}, changes slightly with $\eva$, but stays very near zero. The probe field does affect the atomic variables, but this is effect is extremely small; $\evs{1}{1}$, $\evs{2}{2}$, and $\evs{1}{2}^{(*)}$ are perturbed slightly from their initial steady-state values set by the pump fields, leading to only a small change in $n$. Over most of the $z-t$ region, there is no significant spatial phase variation (see \fig{fig:5La_c_arg}). 

We note again that the instantaneous $n$ calculation ignores terms involving derivatives of $\Emcal \propto \eva$, which might introduce too much error and cause \eq{eq:zi_five-level_index} to be invalid; a more accurate calculation of the index could better explain the observed phase changes not accounted for by $n$.

\section{Conclusion}\label{sec:zi_conclusion}
All of the results from the propagation of a field through three- or five-level atomic media, prepared to give a ZI response, strongly suggest that a single photon does indeed destroy its own ZI. In the three-level case, the strong variation in the instantaneous index with a time-dependent incoming field amplitude shows that the ZI is particularly delicate.

For the five-level atoms, the instantaneous index stays near zero, but phase plots show that the effective index does not, because there is some spatial phase variation, especially along the front of the pulse as it propagates through the medium. However, there is spatial phase change over much less of the $z-t$ region compared to the three-level atoms.

In both cases, we believe that the results are conclusive within the approximations made. However, further study is needed because of these approximations, the semi-classical nature of the models, and the numerical error introduced by the discrete method that was needed to perform the calculations. Still, the spatial phase variation seen in all of the results suggest that it may be difficult or impossible to maintain a ZI, even for a single photon, which diminishes the promise of strong, distant dipole-dipole coupling as well as other applications. Although our calculations were for atomic media, we predict that these conclusions also apply to metamaterials, where the light-matter interaction is still fundamentally between photons and other particles, and where the Kramers-Kronig relations are still valid.

Our findings motivate future work where some of the performed approximations -- especially the mean-field approximation -- are dropped, which should result in more accurate results and further insights in the quantum dynamics of the studied models \cite{javanainen2016}. In addition, tailoring the pulse shape could also be used to minimize the spatial phase change by tailoring the pulse or other parameters. However, the qualitative results presented here are not expected to be altered in this case.

The ultimate goal should be to move beyond these semi-classical models and take a fully quantum approach. In this case the pulse propagation could be solved using a Monte-Carlo treatment, which would more accurately capture the quantum nature of the atom-field interaction.

\section*{Acknowledgments}
We would like to acknowledge the NSF via PHY-1912607 and PHY-2207972.

\appendix
\section{Appendix: Additional Plots for Three-Level Case}
\label{3levelappendix}
\renewcommand{\thefigure}{A\arabic{figure}}
\setcounter{figure}{0}
\renewcommand{\theequation}{A\arabic{equation}}
\setcounter{equation}{0}

\subsection{Contour Plots}
\begin{figure}[H]
    	\includegraphics[trim={0cm 0cm 0cm 0cm},clip,width=\linewidth]{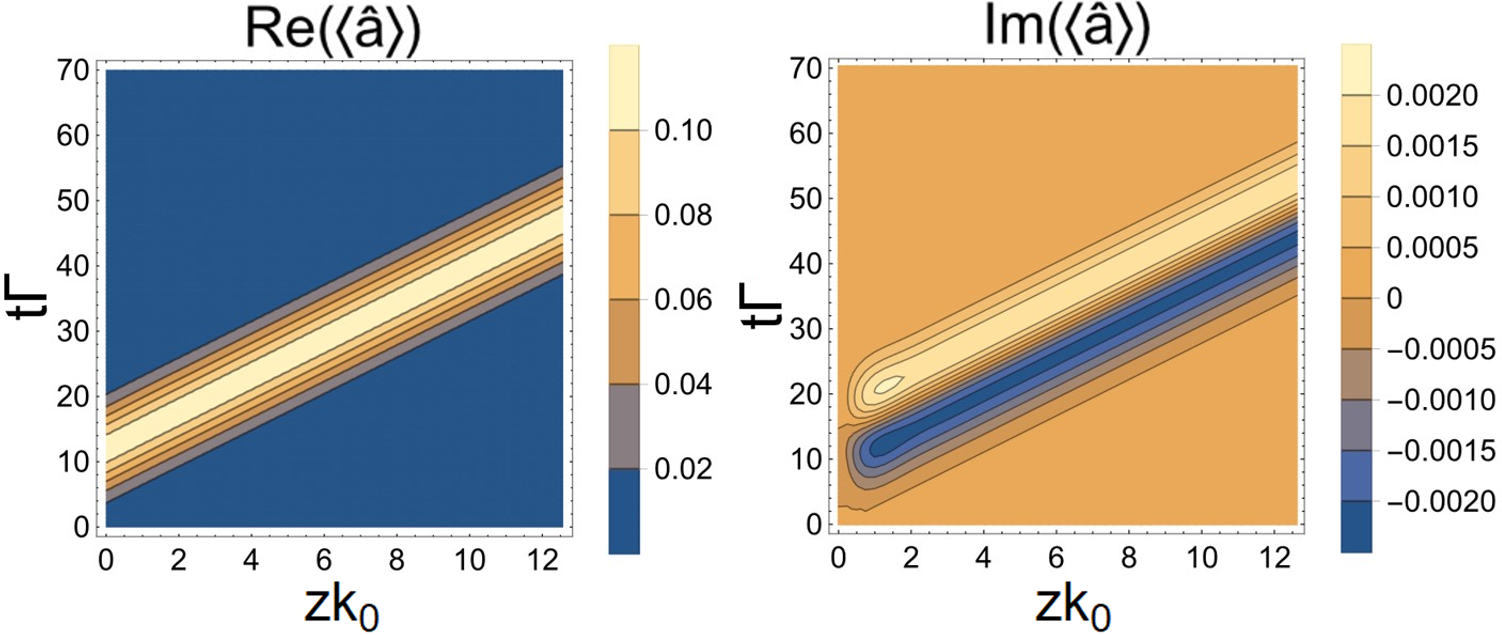}
	\caption{Plot of the pulse $\eva$. It begins purely real at $z=$0 where the pulse enters from vacuum, and the real part is roughly unchanged throughout the propagation. It gains a small imaginary part after $z=0$. Parameters are the same as in \fig{fig:3La_c_arg}.}
     \label{fig:3La_c_a}
\end{figure}

\begin{figure}[H]
\centering
    	\includegraphics[trim={0cm 0cm 0cm 0cm},clip,width=\linewidth]{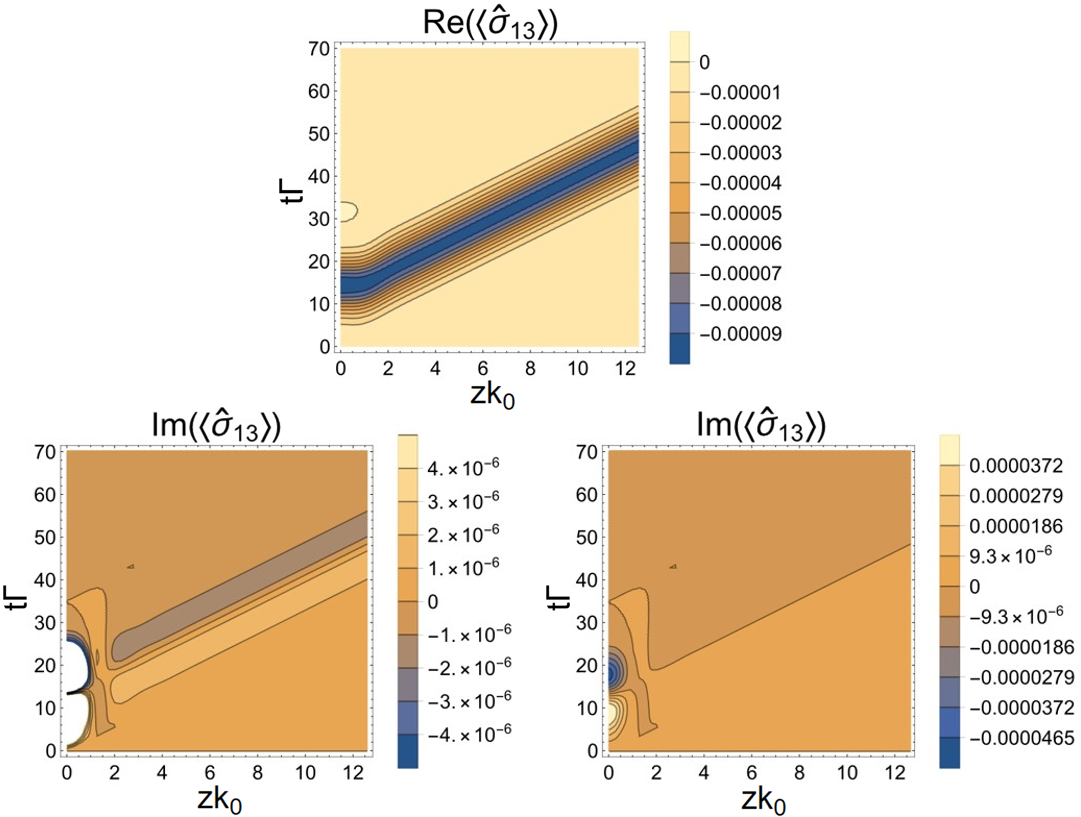}
	\caption{Coherence of the probe-field transition, with two different scales for the imaginary part. $\evs{1}{3}$ is roughly proportional to $\eva$, except for a delay near $z=0$, which is what causes $n$ to vary the most near the boundary. Parameters are the same as in \fig{fig:3La_c_arg}.}
     \label{fig:3La_c_s13}
\end{figure}

\subsection{Time Plots at $z=0$}
\label{CItz=0}

\begin{figure}[H]
\centering
    	\includegraphics[trim={0cm 0cm 0cm 0cm},clip,width=\linewidth]{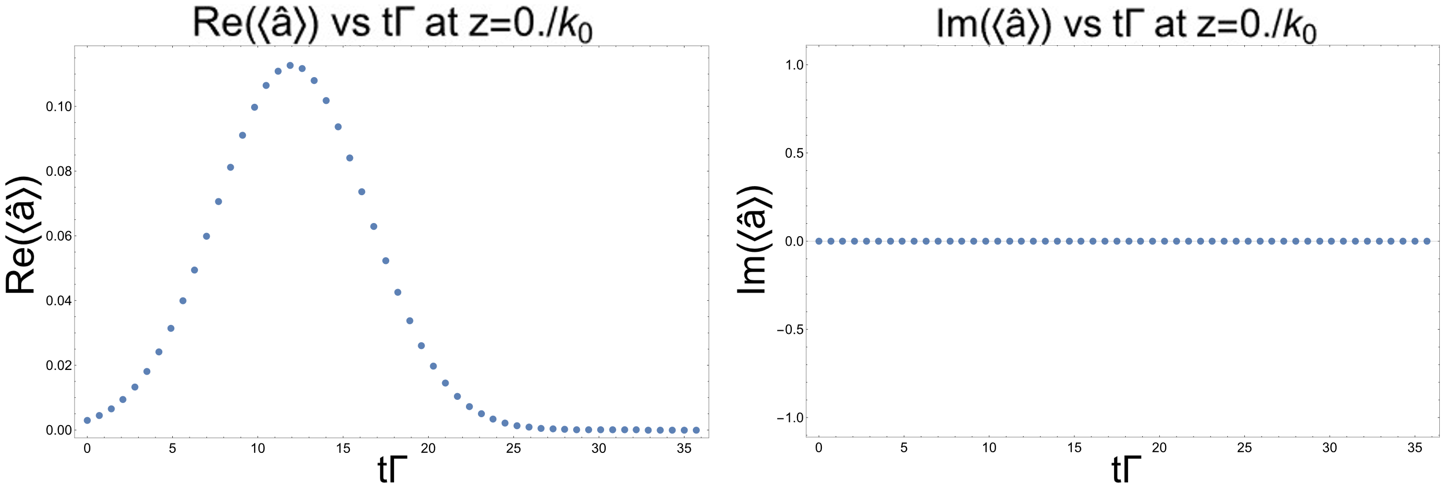}
	\caption{$\eva$ as a function of time at $z=0$. Here, $\eva$ matches the pulse of \eq{eq:zi_pulse}. Parameters are the same as in \fig{fig:3La_c_arg}.}
     \label{fig:3La_t_z=0_a}
\end{figure}

\begin{figure}[H]
\centering
    	\includegraphics[trim={0cm 0cm 0cm 0cm},clip,width=\linewidth]{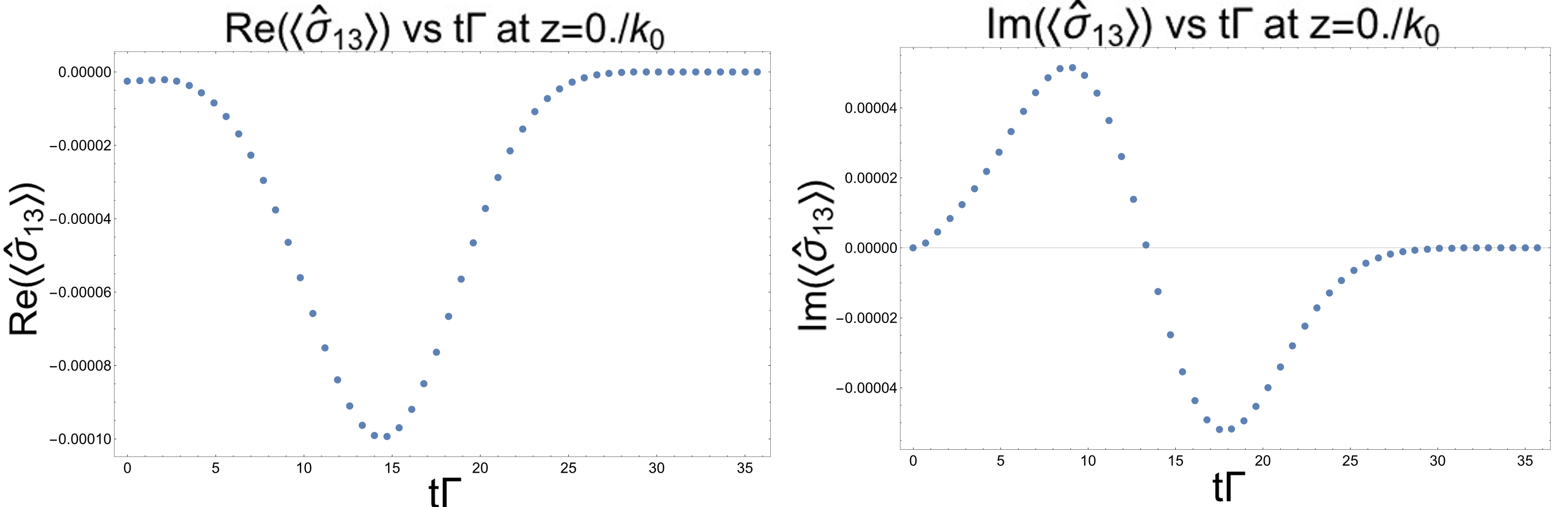}
	\caption{$\evs{1}{3}$ as a function of time at $z=0$. The real part is roughly proportional to $\text{Re}(\eva)$, except it lags for several time steps. Parameters are the same as in \fig{fig:3La_c_arg}.}
     \label{fig:3La_t_z=0_s13}
\end{figure}

\begin{figure}[H]
\centering
    	\includegraphics[trim={0cm 0cm 0cm 0cm},clip,width=\linewidth]{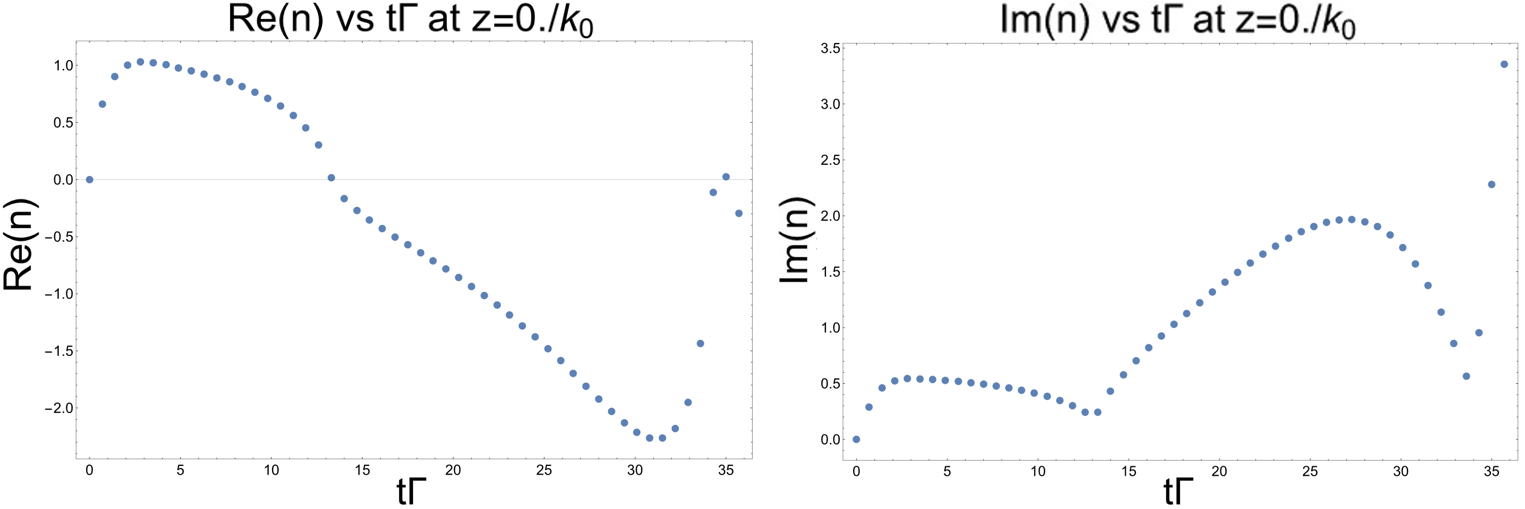}\
	\caption{$n$ as a function of time at $z=0$. Both the real and imaginary parts increase from zero in the first time step, then continue to vary as the pulse enters the medium. In the last few time steps, $\text{Im}(n)$ begins to vary quickly, but this is because the pulse has mostly passed ($\eva$ has fallen to about $10^{-7}$), and $\chi$ in \eq{eq:zi_three-level_index} begins to diverge. This is when the simulation is cut off at this point in $z$. Parameters are the same as in \fig{fig:3La_c_arg}.}
     \label{fig:3La_t_z=0_n}
\end{figure}

\subsection{Time Plots at $z=2\pi/k_0$ (Halfway through Medium)}
\label{CIthalf}

\begin{figure}[H]
\centering
    	\includegraphics[trim={0cm 0cm 0cm 0cm},clip,width=\linewidth]{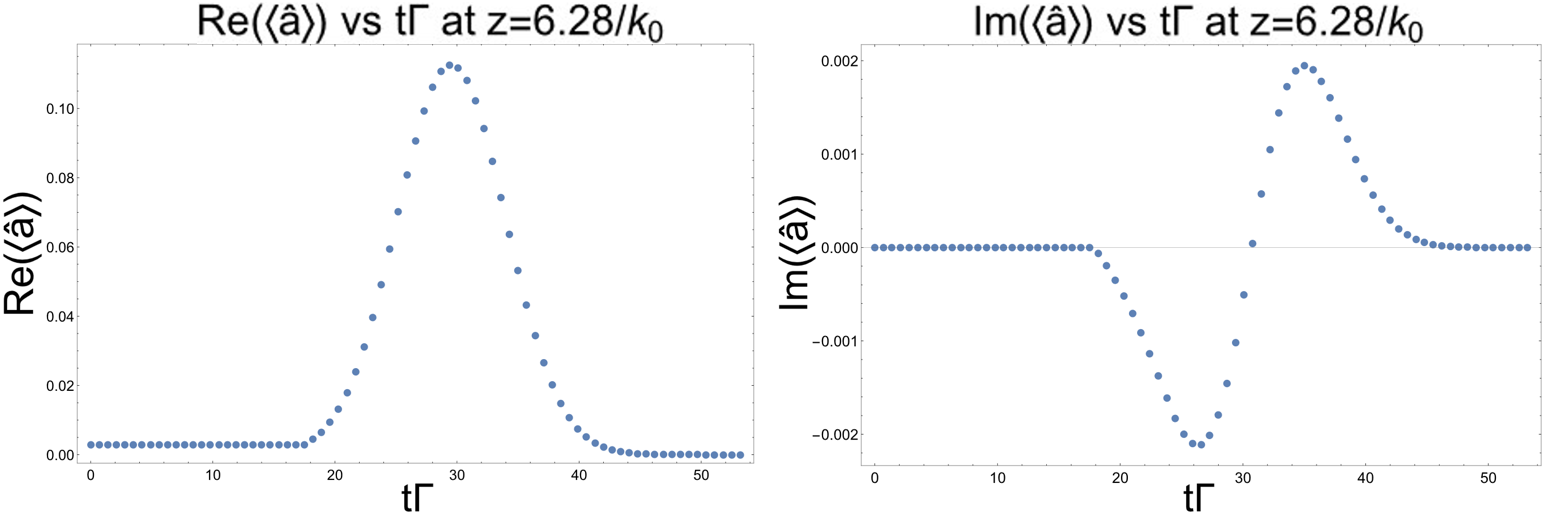}
	\caption{$\eva$ as a function of time, at the point halfway through the medium. The real part is nearly same as at $z=0$, but here an imaginary part has developed. Parameters are the same as in \fig{fig:3La_c_arg}.}
     \label{fig:3La_t_half_a}
\end{figure}

\begin{figure}[H]
\centering
    	\includegraphics[trim={0cm 0cm 0cm 0cm},clip,width=\linewidth]{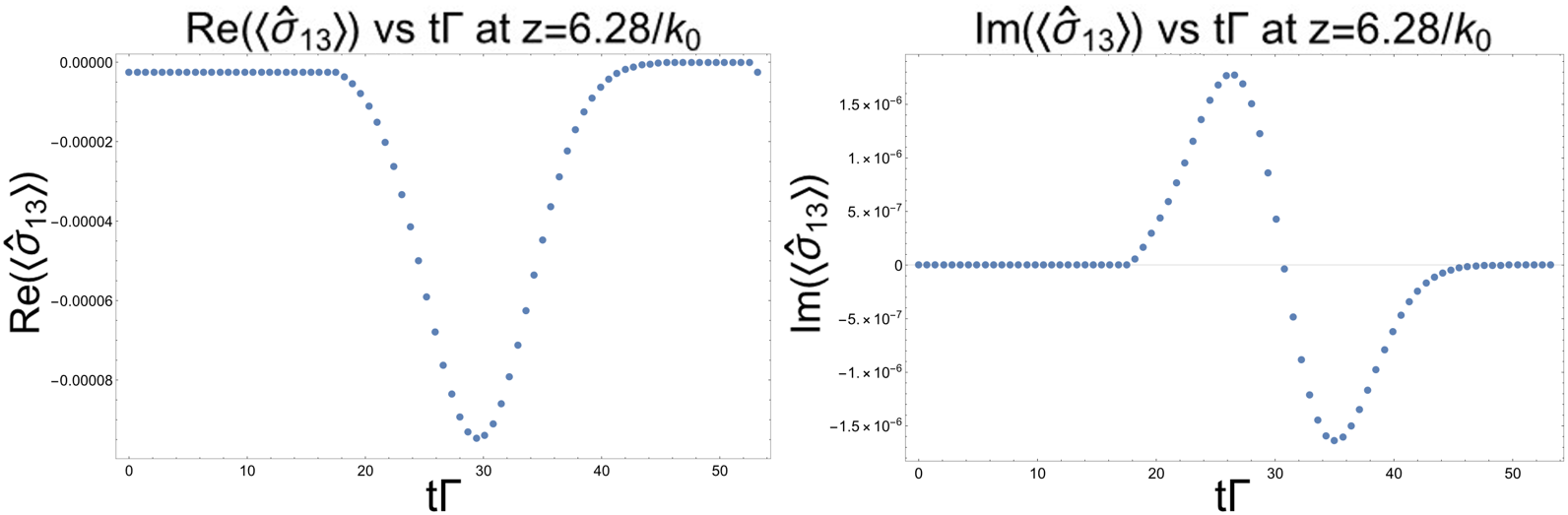}
	\caption{$\evs{1}{3}$ as a function of time, at the point halfway through the medium. Here, it is roughly proportional to $\eva$, so $n$ varies less than at previous points in space. Parameters are the same as in \fig{fig:3La_c_arg}.}
     \label{fig:3La_t_half_s13}
\end{figure}

\begin{figure}[H]
\centering
    	\includegraphics[trim={0cm 0cm 0cm 0cm},clip,width=\linewidth]{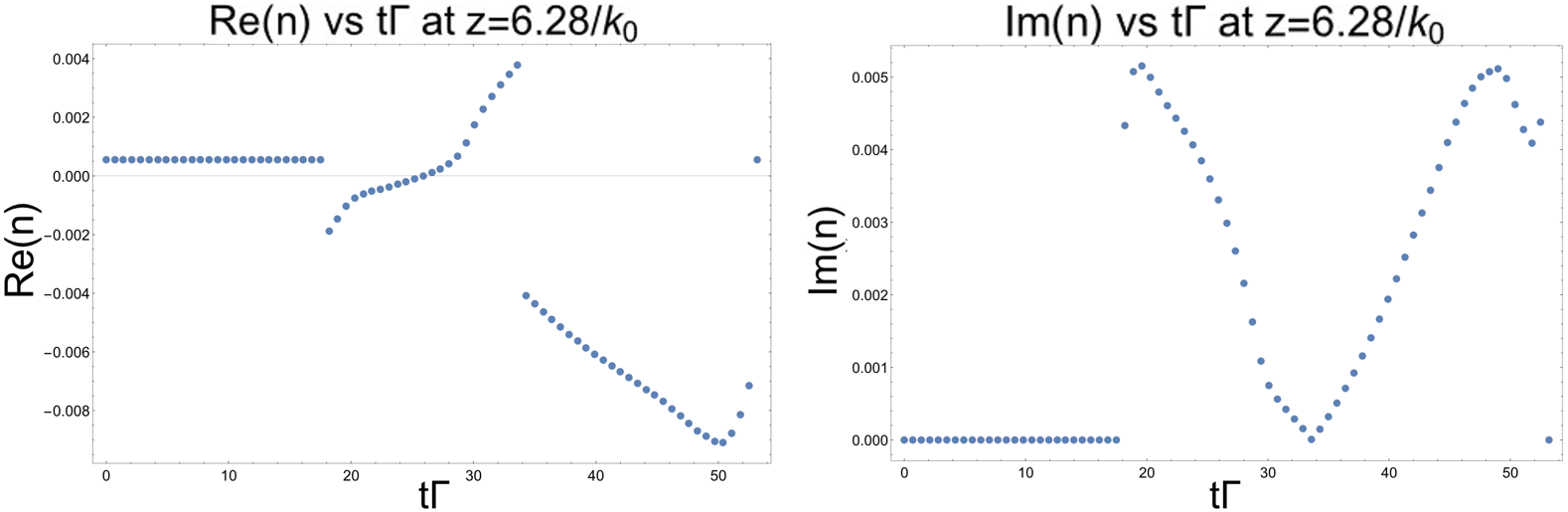}
	\caption{$n$ as a function of time, halfway through the medium. It varies, but stays closer to zero overall than it does nearer to $z=0$. Parameters are the same as in \fig{fig:3La_c_arg}.}
     \label{fig:3La_t_half_n}
\end{figure}

\subsection{Space Plots at $t=0$}
\label{CIst=0}

\begin{figure}[H]
\centering
    	\includegraphics[trim={0cm 0cm 0cm 0cm},clip,width=\linewidth]{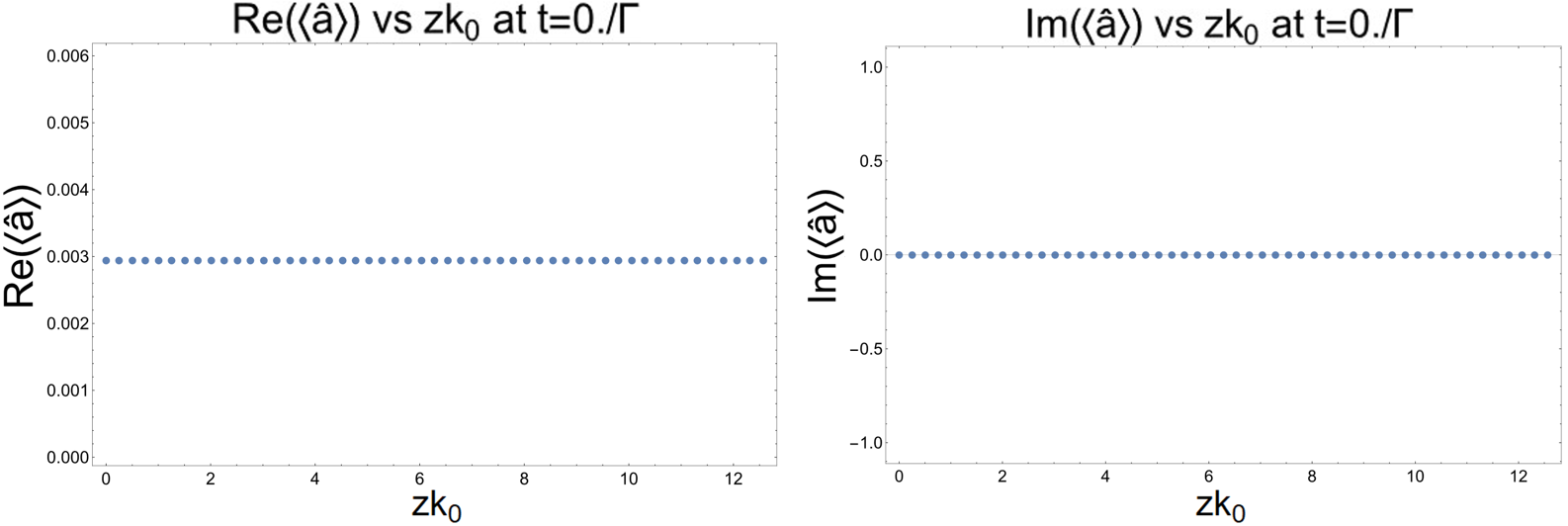}
	\caption{$\eva$ as a function of space at $t=0$. This shows the steady-state value initially prepared throughout the medium to give ZI. Parameters are the same as in \fig{fig:3La_c_arg}.}
     \label{fig:3La_s_t=0_a}
\end{figure}

\begin{figure}[H]
\centering
    	\includegraphics[trim={0cm 0cm 0cm 0cm},clip,width=\linewidth]{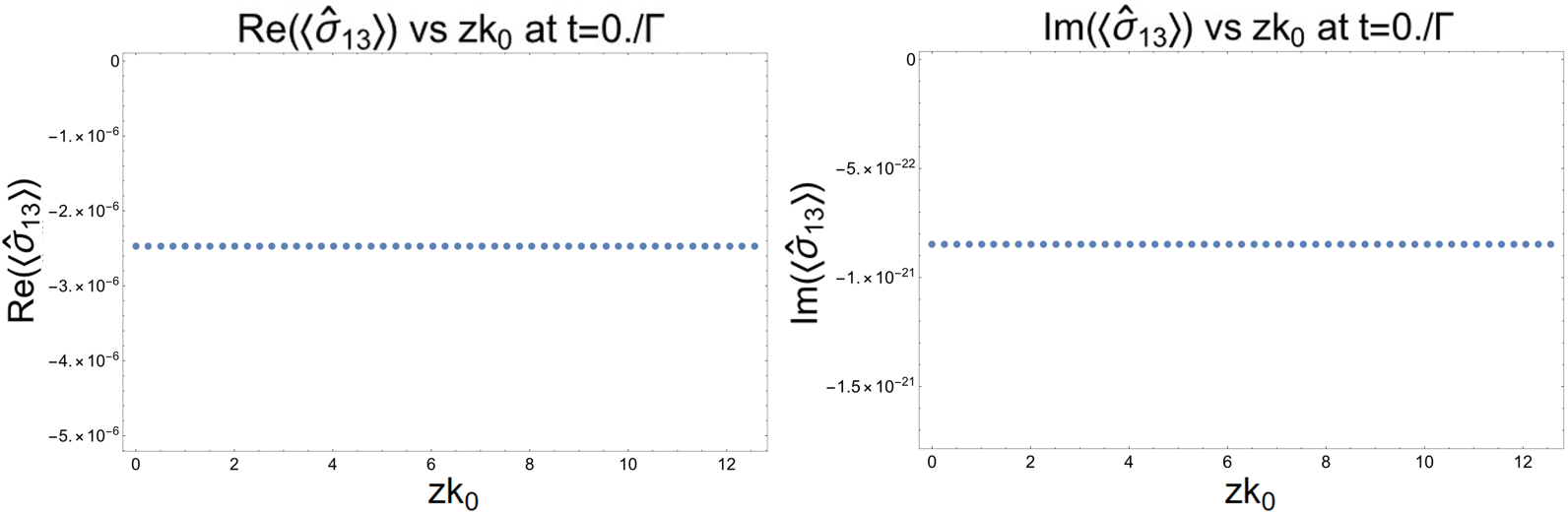}
	\caption{$\evs{1}{3}$ as a function of space at $t=0$. This shows the steady-state value initially prepared throughout the medium to give ZI. Parameters are the same as in \fig{fig:3La_c_arg}.}
     \label{fig:3La_s_t=0_s13}
\end{figure}

\begin{figure}[H]
\centering
    	\includegraphics[trim={0cm 0cm 0cm 0cm},clip,width=\linewidth]{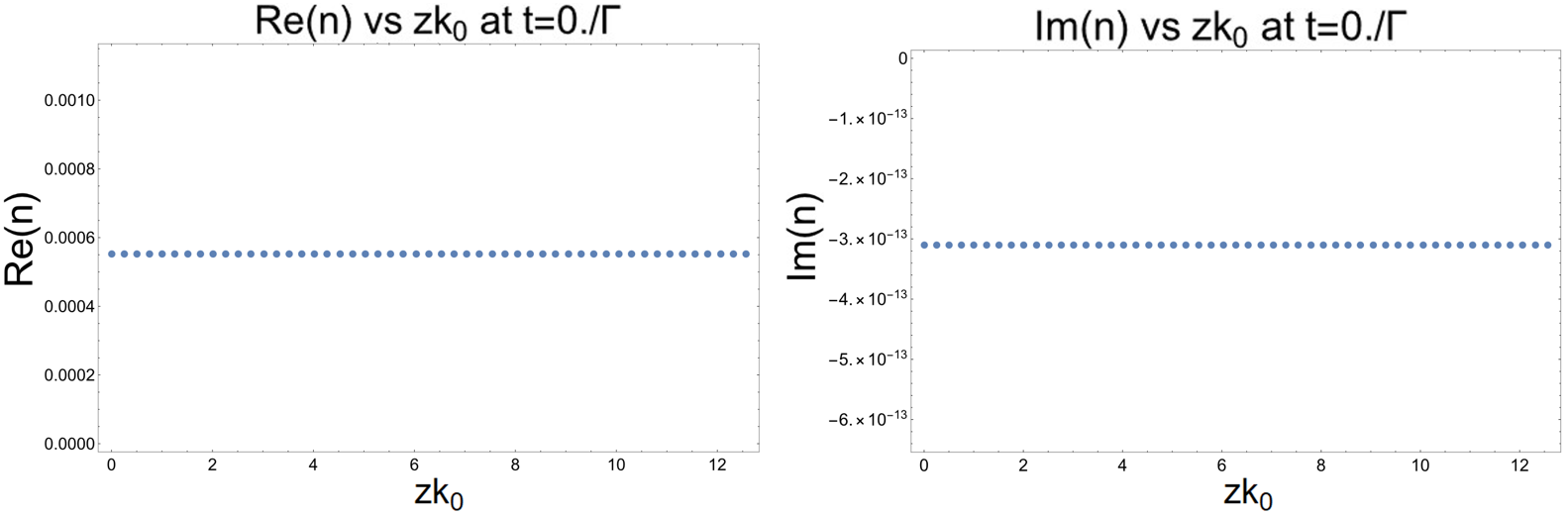}
	\caption{$n$ as a function of space at $t=0$. This shows the near-zero initial value initially prepared throughout the medium. Parameters are the same as in \fig{fig:3La_c_arg}.}
     \label{fig:3La_s_t=0_n}
\end{figure}

\subsection{Space Plots when Field Peak is Halfway through Medium}
\label{CIshalfway}

\begin{figure}[H]
\centering
    	\includegraphics[trim={0cm 0cm 0cm 0cm},clip,width=\linewidth]{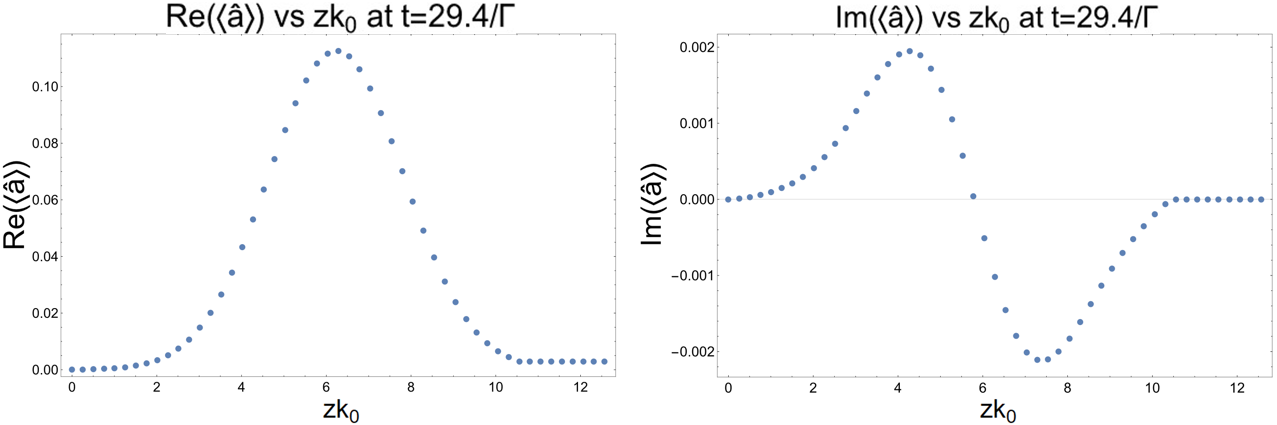}
	\caption{$\eva$ as a function of space, when its peak is halfway through the medium. Parameters are the same as in \fig{fig:3La_c_arg}.}
     \label{fig:3La_s_peakhalfway_a}
\end{figure}

\begin{figure}[H]
\centering
    	\includegraphics[trim={0cm 0cm 0cm 0cm},clip,width=\linewidth]{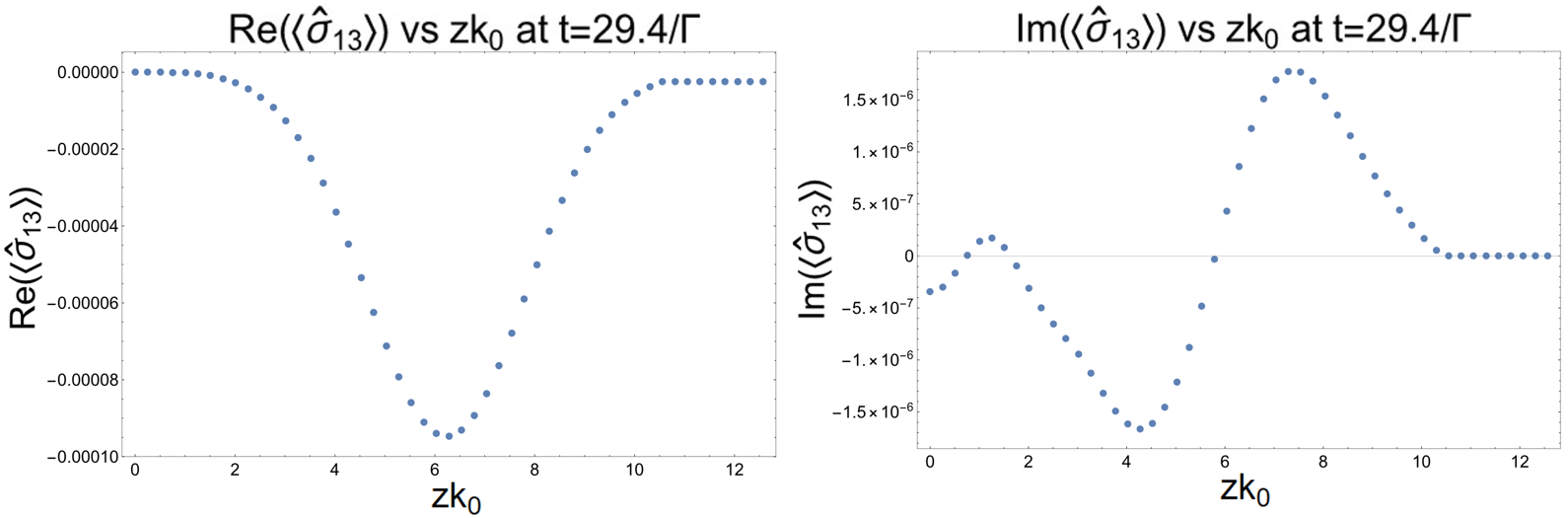}
	\caption{$\evs{1}{3}$ as a function of space, when the peak of $\eva$ is halfway through the medium. It is roughly proportional to $\eva$, except for at low $z$, especially below $zk_0=4$, where $n$ varies the most. Parameters are the same as in \fig{fig:3La_c_arg}.}
     \label{fig:3La_s_peakhalfway_s13}
\end{figure}

\begin{figure}[H]
\centering
    	\includegraphics[trim={0cm 0cm 0cm 0cm},clip,width=\linewidth]{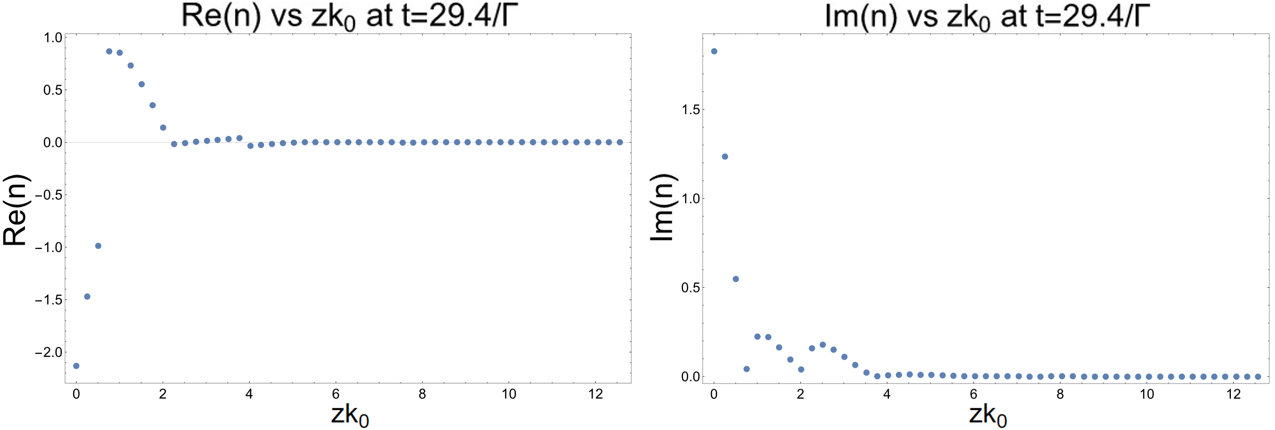}
	\caption{$n$ as a function of space, when the peak of $\eva$ is halfway through the medium. Both parts of $n$ still vary away from zero the most near $z=0$ where the pulse is entering, but stay closer to zero further inside the medium. Parameters are the same as in \fig{fig:3La_c_arg}.}
     \label{fig:3La_s_peakhalfway_n}
\end{figure}

\section{Bloch Equations: Five-Level Case}
\label{five-level_bloch}
The Bloch equations derived from the Lindblad master equation with \eq{eq:zi_five-level_hamiltonian} are:
\begin{equation*}
    \begin{split}
        \pdv{\evs{1}{2}}{t} &\approx \;ig^*\evad\evs{1}{4} - ig_B\eva\evs{2}{3}^* - i\Omega_1\evs{2}{5}^* \\
        &+ i\Omega_2^*\evs{1}{5}, \\
        \pdv{\evs{1}{3}}{t} &\approx \; \left[-i(\delta-\Delta)-\Gamma_{31}/2\right]\evs{1}{3} \\
        &- ig_B\eva\left(\evs{3}{3}-\evs{1}{1}\right)- i\Omega_1\evs{3}{5}^* \\
        &+ i \Omega_3^*\evs{1}{4}, \\
        \pdv{\evs{1}{4}}{t} &\approx \;\left(i\Delta-\Gamma_{42}/2-\Gamma_{43}/2\right)\evs{1}{4} + ig\eva\evs{1}{2} \\
        &- ig_B\eva\evs{3}{4} - i\Omega_1\evs{4}{5}^* + i\Omega_3\evs{1}{3}, \\
        \pdv{\evs{1}{5}}{t} &\approx \;-\left(\Gamma_{51}+\Gamma_{52}\right)/2\evs{1}{5} - ig_B\eva\evs{3}{5} \\
        &- i\Omega_1\left(\evs{5}{5}-\evs{1}{1}\right) + i\Omega_2\evs{1}{2}, \\
        \pdv{\evs{2}{3}}{t} &\approx \;\left[-i(\delta-\Delta)-\Gamma_{31}/2\right]\evs{2}{3} - ig\eva\evs{4}{3} \\
        &+ ig_B\eva\evs{1}{2}^* - i\Omega_2\evs{3}{5} + i\Omega_3^*\evs{2}{4},
        \end{split}
        \end{equation*}
        \begin{equation*}
            \begin{split}
        \pdv{\evs{2}{4}}{t} &\approx \;\left(i\Delta-\Gamma_{42}/2-\Gamma_{43}/2\right)\evs{2}{4} \\
        &- ig\eva\left(\evs{4}{4}-\evs{2}{2}\right) - i\Omega_2\evs{4}{5}^* + i\Omega_3\evs{2}{3}, \\
        \pdv{\evs{2}{5}}{t} &\approx \;-\left(\Gamma_{51}+\Gamma_{52}\right)/2\evs{2}{5} - ig\eva\evs{4}{5} \\
        &+ i\Omega_1\evs{1}{2}^* - i\Omega_2\left(\evs{5}{5}-\evs{2}{2}\right), \\
        \pdv{\evs{3}{4}}{t} &\approx \;\left(i\delta-\Gamma_{31}/2-\Gamma_{42}/2-\Gamma_{43}/2\right)\evs{3}{4} \\
        &+ ig\eva\evs{2}{3}^* - ig_B\evad\evs{1}{4} \\
        &- i\Omega_3\left(\evs{4}{4}-\evs{3}{3}\right), \\
        \pdv{\evs{3}{5}}{t} &\approx \;\left[i(\delta-\Delta)-\Gamma_{31}/2-\Gamma_{51}/2-\Gamma_{52}/2\right]\evs{3}{5} \\
        &- ig_B^*\evad\evs{1}{5} + i\Omega_1\evs{1}{3}^* \\
        & + i\Omega_2\evs{2}{3}^* - i\Omega_3\evs{4}{5}, \\
        \pdv{\evs{4}{5}}{t} &\approx \;(-i\Delta-\Gamma_{42}/2-\Gamma_{43}/2-\Gamma_{51}/2-\Gamma_{52}/2)\evs{4}{5} \\
        &- ig^*\evad\evs{3}{5} + i\Omega_1\evs{1}{4}^* - i\Omega_2\evs{2}{4}^*, \\
        \pdv{\evs{1}{1}}{t} &\approx \;-ig_B\eva\evs{1}{3}^* + ig_B^*\evad\evs{1}{3} - i\Omega_1\evs{1}{5}^* \\
        &+ i\Omega_1^*\evs{1}{5} + \Gamma_{31}\evs{3}{3} + \Gamma_{51}\evs{5}{5}, \\
        \pdv{\evs{2}{2}}{t} &\approx \;-ig\eva\evs{2}{4}^* + ig^*\evad\evs{2}{4} - i\Omega_2\evs{2}{5}^* \\
        &+ i\Omega_2^*\evs{2}{5} + \Gamma_{42}\evs{4}{4} + \Gamma_{52}\evs{5}{5}, \\
        \pdv{\evs{3}{3}}{t} &\approx \;ig_B\eva\evs{1}{3}^* - ig_B^*\evad\evs{1}{3} - i\Omega_3\evs{3}{4}^* \\
        &+ i\Omega_3^*\evs{3}{4} - \Gamma_{31}\evs{3}{3} + \Gamma_{43}\evs{4}{4},
    \end{split}
\end{equation*}
\begin{equation}\label{eq:zi_five-level_bloch}
    \begin{split}
        \pdv{\evs{4}{4}}{t} \approx& \;ig\eva\evs{2}{4}^* - ig^*\evad\evs{2}{4} + i\Omega_3\evs{3}{4}^* \\
        &- i\Omega_3^*\evs{3}{4} - (\Gamma_{42}+\Gamma_{43})\evs{4}{4}, \\
        \pdv{\evs{5}{5}}{t} =& \;i\Omega_1\evs{1}{5}^* - i\Omega_1^*\evs{1}{5} + i\Omega_2\evs{2}{5}^* \\
        &- i\Omega_2^*\evs{2}{5} - (\Gamma_{51}+\Gamma_{52})\evs{5}{5}.
    \end{split}
\end{equation}

\section{Additional Plots: Five-Level Case}
\renewcommand{\thefigure}{B\arabic{figure}}
\setcounter{figure}{0}
\renewcommand{\theequation}{B\arabic{equation}}
\setcounter{equation}{0}

\subsection{Contour Plots}

\begin{figure}[H]
\centering
    	\includegraphics[trim={0cm 0cm 0cm 0cm},clip,width=\linewidth]{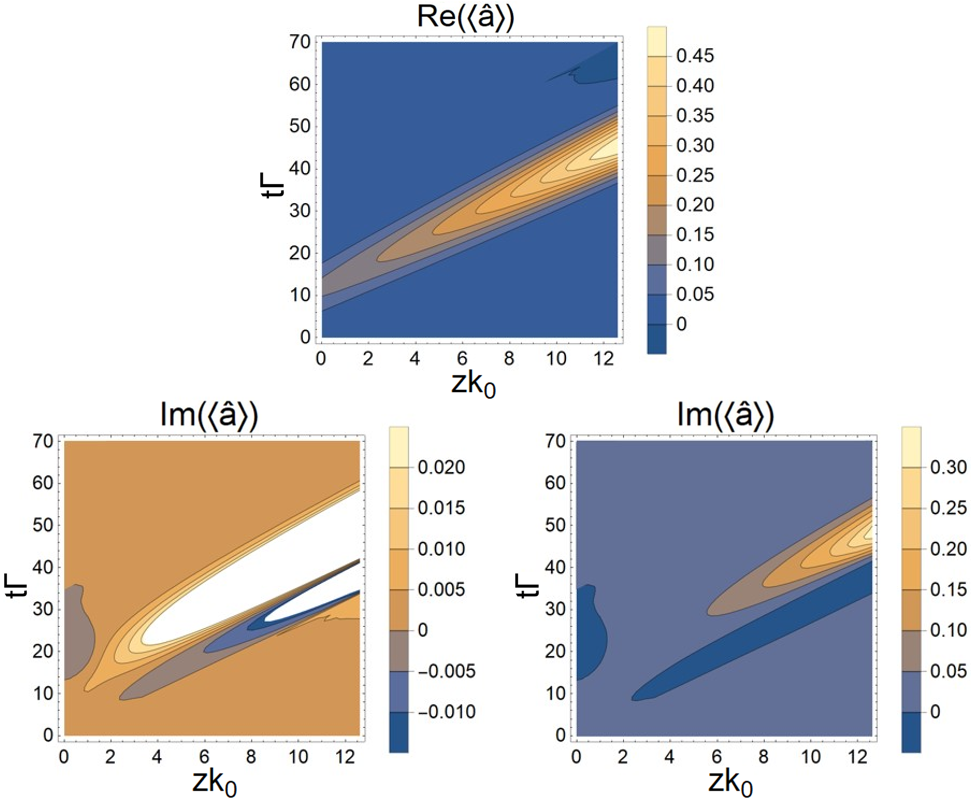}
	\caption{$\eva$ in space and time, with two different scales for the imaginary part. At $z=0$, $\eva$ is purely real, but gains an imaginary part upon entering the medium. Both parts widen and grow as the pulse travels further into the medium.}
     \label{fig:5La_c_a}
\end{figure}

\subsection{Time Plots at $z=0$ and at $z=z_f$}
\label{5Lat}

\begin{figure}[H]
    \centering
    \includegraphics[trim={0cm 0cm 0cm 0cm},clip,width=0.5\linewidth]{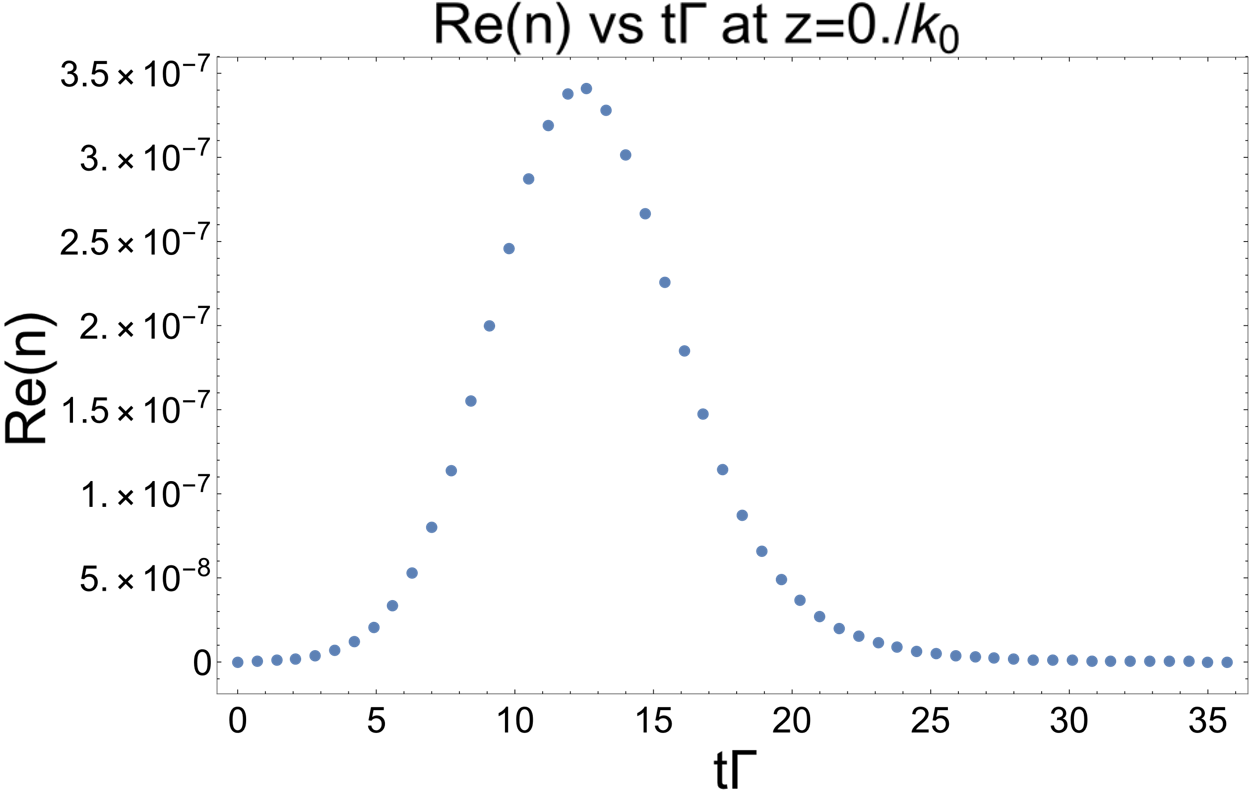}
	\caption{$\text{Re}(n)$ as a function of time at $z=0$. It increases very slightly with $\eva$, then returns to its initial value after the pulse passes.}
     \label{fig:5La_t_z=0_n}
\end{figure}

\begin{figure}[H]
    \centering
    \includegraphics[trim={0cm 0cm 0cm 0cm},clip,width=0.5\linewidth]{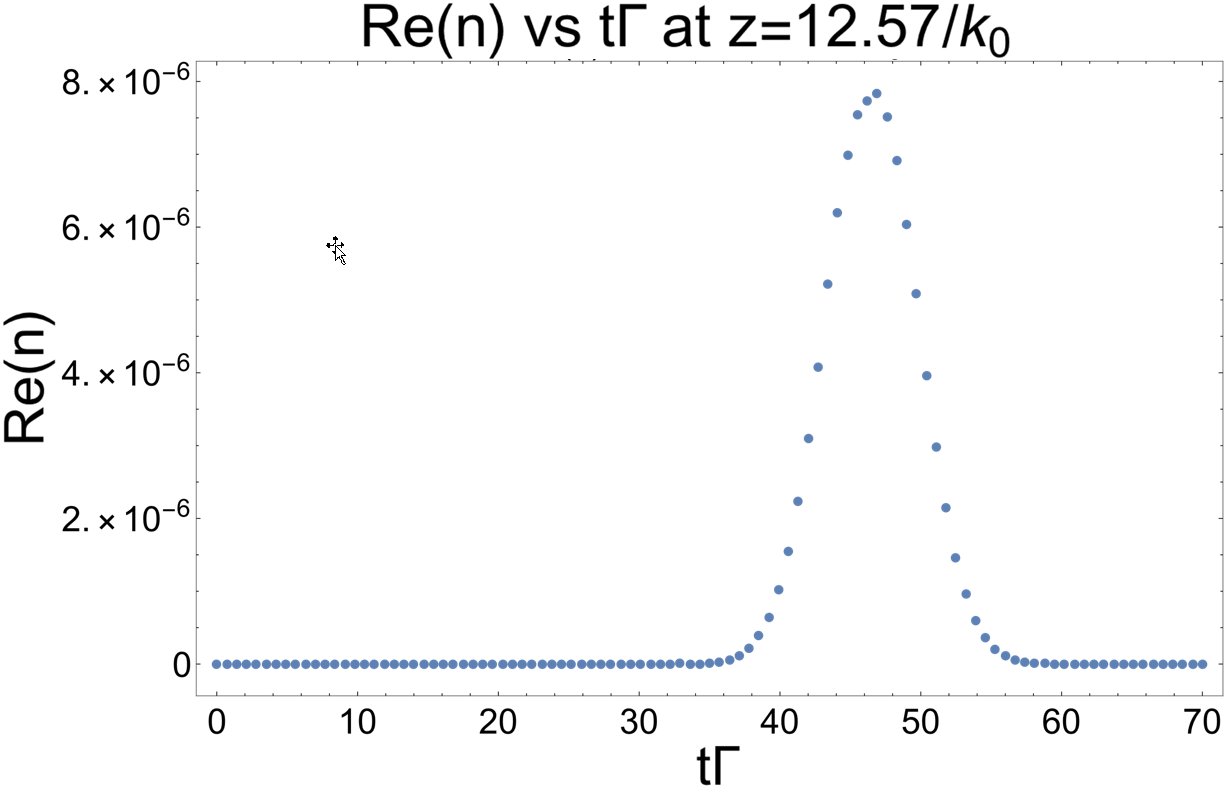}
	\caption{$\text{Re}(n)$ as a function of time at $z=z_f$. It increases slightly more than at $z=0$ due to a growing $\eva$ further into the medium, but is still very near zero.}
     \label{fig:5La_t_z=zf_n}
\end{figure}

\begin{figure}[H]
\centering
    	\includegraphics[trim={0cm 0cm 0cm 0cm},clip,width=\linewidth]{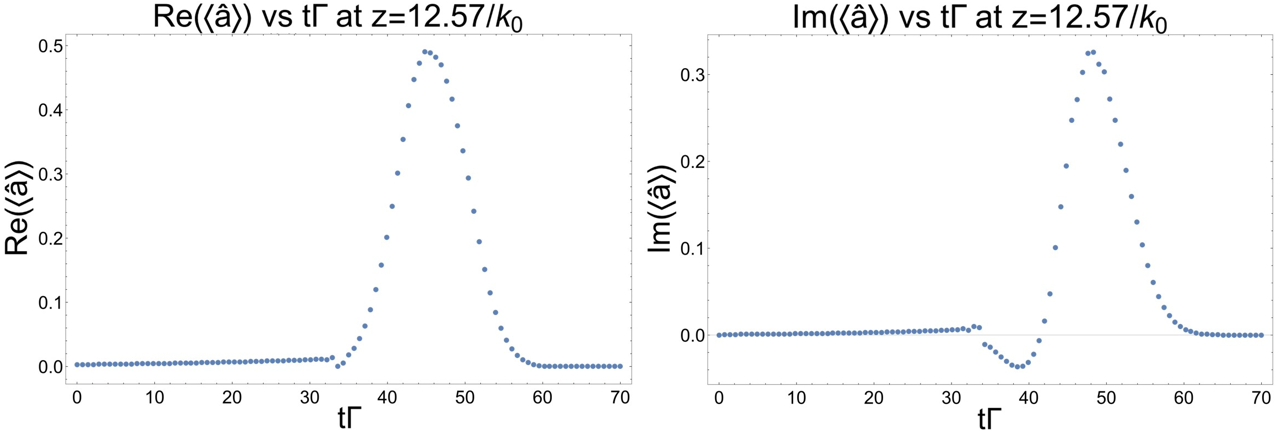}
	\caption{$\eva$ as a function of time at $z=z_f$. (At $z=0$, $\eva$ is identical to \fig{fig:3La_t_z=0_a}, where it is given by \eq{eq:zi_pulse}.) There is an apparent discontinuity or rapid change at $t\Gamma=34$, which can be traced back to $z=0$, $t=0$, where the pulse first begins interacting with the medium. This corresponds to the sharp spatial phase change just above the red line in \fig{fig:5La_c_arg}.}
     \label{fig:5La_t_z=zf_a}
\end{figure}

\bibliography{zipropagation}

\end{document}